\documentclass[aps,prd,preprint,superscriptaddress,nofootinbib]{revtex4-1}
\usepackage{amsmath}
\usepackage{amssymb}
\usepackage{graphicx}% Include figure files
\usepackage{dcolumn}% Align table columns on decimal point
\usepackage{bm}% bold math
\usepackage{hyperref}% add hypertext capabilities
\usepackage{epsfig}
\usepackage{mathrsfs}
\usepackage{slashed}
\usepackage{ulem}

\newcommand{\nbar}{\bar{n}}
\newcommand{\slnbar}{\slashed{\bar{n}}}

\newcommand{\eps}{\varepsilon}

\begin{document}

% Use the \preprint command to place your local institutional report
% number in the upper righthand corner of the title page in preprint mode.
% Multiple \preprint commands are allowed.
% Use the 'preprintnumbers' class option to override journal defaults
% to display numbers if necessary
\preprint{MIT-CTP/5967}

%%%%%%%%%%%%%%%%%%%%%%%%%%%%%%%%%%%%%%%%%%%%%%%%%%%%%%%%%%%%%%
%Title of paper
\title{Hidden self-energy contributions of collinear functions in SCET}
%%%%%%%%%%%%%%%%%%%%%%%%%%%%%%%%%%%%%%%%%%%%%%%%%%%%%%%%%%%%%%

% repeat the \author .. \affiliation  etc. as needed
% \email, \thanks, \homepage, \altaffiliation all apply to the current
% author. Explanatory text should go in the []'s, actual e-mail
% address or url should go in the {}'s for \email and \homepage.
% Please use the appropriate macro foreach each type of information

% \affiliation command applies to all authors since the last
% \affiliation command. The \affiliation command should follow the
% other information
% \affiliation can be followed by \email, \homepage, \thanks as well.

%%%%%%%%%%%%%%%%%%%%%%%%%%%%%%%%%%%%%%%%%%%%%%%%%%%%%%%%%%%%%%
\author{Geoffrey~T.~Bodwin}
\email[]{Contact author: gtb@anl.gov}
\affiliation{High Energy Physics Division, Argonne National Laboratory,
Argonne, Illinois 60439, USA}
\author{June-Haak~Ee}
\email[]{Contact author: jhee@mit.edu}
\affiliation{Center for Theoretical Physics – a Leinweber Institute, Massachusetts Institute of Technology, Cambridge, MA 02139, USA}
\author{Daekyoung Kang}
\email[]{Contact author: dkang@fudan.edu.cn}
\affiliation{Key Laboratory of Nuclear Physics and Ion-beam Application (MOE) and Institute of Modern Physics, Fudan University, Shanghai 200433, China}
\affiliation{Department of Physics, Korea University, Seoul 02841, Korea} 
\author{Xiang-Peng~Wang}
\email[]{Contact author: xpwang@ccnu.edu.cn}
\affiliation{Institute of Particle Physics and Key Laboratory of Quark and Lepton Physics (MOE), Central
China Normal University, Wuhan, Hubei 430079, China} 
%%%%%%%%%%%%%%%%%%%%%%%%%%%%%%%%%%%%%%%%%%%%%%%%%%%%%%%%%%%%%%

%\email[]{Your e-mail address}
%\homepage[]{Your web page}
%\thanks{}
%\altaffiliation{}
%\affiliation{}

%Collaboration name if desired (requires use of superscriptaddress
%option in \documentclass). \noaffiliation is required (may also be
%used with the \author command).
%\collaboration can be followed by \email, \homepage, \thanks as well.
%\collaboration{}
%\noaffiliation

\date{\today}

%%%%%%%%%%%%%%%%%%%%%%%%%%%%%%%%%%%%%%%%%%%%%%%%%%%%%%%%%%%%%%
\begin{abstract}
The LSZ reduction formula requires one to identify and amputate complete
propagators on external legs of a Green's function and to evaluate
complete two-point functions in the mass-shell limit.  Motivated by
these requirements, we analyze quark self-energy contributions on
external legs in soft-collinear effective theory (SCET).  We examine an
operator basis that follows directly from full quantum chromodynamics
(QCD) (upon application of the SCET equations of motion to express small
Dirac components in terms of large Dirac components). We find that, for
this basis, the self-energy contributions can be identified from their
diagrammatic topologies, as in full QCD.  However, for an alternative
operator basis that is obtained from the direct-QCD basis by an
application of Wilson-line identities, interactions are shifted from a
covariant derivative to a Wilson line.  Consequently, some self-energy
contributions are hidden in diagrams involving Wilson lines, making
their identification subtle. We find that the hidden self-energy
    contributions to the two-point function are ill-defined in the
    mass-shell limit, making their computation problematic.  We
introduce a generalization of the LSZ formula that allows one to make
different choices for the complete propagator and that compensates for
those choices through the factor that arises from the on-shell residue
of the two-point function.  We use this generalization to explore, in
    both SCET operator bases, various options for using the LSZ formula
    to construct the $S$-matrix.
\end{abstract}
%%%%%%%%%%%%%%%%%%%%%%%%%%%%%%%%%%%%%%%%%%%%%%%%%%%%%%%%%%%%%%

% insert suggested keywords - APS authors don't need to do this
%\keywords{}

%%%%%%%%%%%%%%%%%%%%%%%%%%%%%%%%%%%%%%%%%%%%%%%%%%%%%%%%%%%%%%
%\maketitle must follow title, authors, abstract, and keywords
\maketitle
%%%%%%%%%%%%%%%%%%%%%%%%%%%%%%%%%%%%%%%%%%%%%%%%%%%%%%%%%%%%%%

% body of paper here - Use proper section commands
% References should be done using the \cite, \ref, and \label commands

% Put \label in argument of \section for cross-referencing
%\section{\label{}}

%%%%%%%%%%%%%%%%%%%%%%%%%%%%%%%%%%%%%%%%%%%%%%%%%%%%%%%%%%%%%%

\section{Introduction}

In this paper, we discuss a difficulty that arises in the construction
    of the $S$-matrix in soft-collinear effective theory (SCET)
\cite{Bauer:2000ew, Bauer:2000yr, Bauer:2001ct, Bauer:2001yt,
  Bauer:2002nz} when one chooses certain operator bases for collinear
functions.  The difficulty has to do with the identification of
self-energy contributions on external legs. It first appears at
subleading power in the SCET expansion \cite{Beneke:2002ph,
  Beneke:2002ni,Pirjol:2002km,Bauer:2003mga}.

In the perturbative application of the LSZ reduction formula for the
construction of the $S$-matrix \cite{Lehmann:1954rq}, one amputates
complete propagators (which include self-energy diagrams) on external
legs of Green's functions and multiplies by a factor $\sqrt{R}$ for each
external leg, where $R$ is the residue of the two-point function of the
external field at the single-particle pole.\footnote{See, for example,
Sec.~7.2 of Ref.~\cite{Peskin:1995ev}.}  The complete propagators in
full QCD consist, of course, of single-particle-irreducible (1PI)
self-energy diagrams, alternating with free-particle propagators.  In
full QCD, it is simple to identify the self-energy contributions on
external legs from their diagrammatic topologies. However, as we will
see, the topological identification of self-energy contributions fails
when certain bases of operators are used to construct collinear
functions in SCET, and self-energy contributions are hidden in diagrams
that involve Wilson lines.  These diagrams do not have the topology of a
self-energy contribution.  A further difficulty in identifying
self-energy contributions is that Wilson-line diagrams can contain both
a self-energy contribution and a non-self-energy contribution. In
    addition, we find that the hidden self-energy contributions to the
    two-point function are ill-defined in the mass-shell limit, making
    their computation as isolated quantities problematic.

These difficulties are absent when one uses a collinear-function
operator basis that follows directly from QCD upon application of the
SCET equations of motion to eliminate the small Dirac components (the
direct-QCD basis).\footnote{For example, the authors of
Ref.~\cite{vanBijleveld:2025ekz} investigated the jet function that
appears in the massive fermion form factor in QED. In contrast with the
standard SCET approach, the approach that they used defines the jet
function in terms of the full QED fields. Hidden self-energy
contributions do not appear in this formalism.}  However, hidden
self-energy contributions can occur when one uses Wilson-line identities
to construct a new basis (the modified basis).  This basis can also be
derived through symmetry arguments alone, without reference to the
underlying UV theory (QCD).  It is possible that the difficulty that we
describe in this paper also arises when one uses symmetry arguments to
derive operator bases in other effective field theories.

We construct a generalization of the LSZ formula that allows one to make
various choices for the complete propagator on external legs that is to
be amputated. In the generalized formula, a change in the choice of the
complete propagator is compensated for by a change in $R$, leaving the
$S$-matrix invariant.  Using this generalized formula, we explore
various options for the construction of the $S$-matrix, in both the
direct-QCD SCET basis and the modified SCET basis. We find that a
procedure in which all of the Wilson-line contributions on external legs
are amputated and, instead, appear in the two-point function, eliminates
the need to identify the hidden self-energy contributions
separately. This procedure is equivalent to the use of the light-cone
gauge.  However, it resolves ambiguities in the light-cone-gauge
    calculation that arise from the denominators of the light-cone-gauge
    gluon-propagator polarization sum.

The remainder of this paper is organized as follows. In
Sec.~\ref{sec:Preliminaries}, we establish our notation and remind the
reader of the power counting for collinear momenta and fields in SCET.
In Sec.~\ref{sec:bases}, we present two collinear-operator bases: one,
    the direct-QCD basis, that is derived, using the SCET equations of
motion, directly from QCD and one, the modified basis, that
follows from an application of a Wilson-line identity.  In
Section~\ref{sec:self-energy-calcs}, we examine quark self-energy
    contributions on external legs in full QCD and in the two SCET
    operator bases.  Here, we find that the calculations in the operator
    basis that is derived by making use of a Wilson-line identity
    require contributions from diagrams involving Wilson lines in order
    to reproduce the self-energy contributions of full QCD at subleading
    power in the SCET expansion. In Sec.~\ref{sec:complete-LSZ}, we
    present a generalization of the complete LSZ formula and use it to
    discuss various options for the construction of the $S$-matrix in
    the direct-QCD and modified bases. We summarize our results in
    Sec.~\ref{sec:summary}. The computation of the hidden self-energy
    contribution to $R$ and computations in the light-cone gauge are
    discussed in Appendices~\ref{app:delta-R} and
    \ref{app:light-cone-gauge}, respectively.

\section{Preliminaries}
\label{sec:Preliminaries}%

We decompose an arbitrary vector in terms of the two light-like vectors, 
$n$ and $\bar{n}$, as follows:
%---------------
\begin{eqnarray}
%---------------
r^\mu = r^- \frac{\bar{n}^\mu}{2}
+r^+ \frac{n^\mu}{2}
+r_\perp^\mu,
%---------------
\end{eqnarray}
%---------------
where 
%---------------
\begin{eqnarray}
\label{eq:light-front-coordinate-convention}
%---------------
r^+ = \bar{n}\cdot r,
\quad
r^- = n\cdot r,
\quad
r_\perp^\mu = r^\mu 
-r^- \frac{\bar{n}^\mu}{2}
-r^+ \frac{n^\mu}{2},
%---------------
\end{eqnarray}
%---------------
with $n^2 = \bar{n}^2 = 0$ and $n\cdot \bar{n}=2$. 
$n$ and $\bar{n}$ are the light-like unit vectors 
along the $z$ axis:
%---------------
\begin{equation}
\label{eq:n-nbar-def}
%---------------
n^\mu = (1,0,0,1),
\quad
\bar{n}^\mu = (1,0,0,-1).
%---------------
\end{equation}
%---------------
The perpendicular momentum $r_\perp^\mu$ satisfies 
$n\cdot r_\perp = \bar{n}\cdot r_\perp =0$. 

We consider the case of a nonzero quark mass $m$ and assume, for
simplicity, that $m \gg \Lambda_{\rm QCD}$. In this case, the SCET
scaling parameter is defined as $\lambda = m/Q$, where $Q$ is the hard
scale of the process. (In the massless case, the SCET scaling parameter
is $\lambda=\Lambda_{\rm QCD}/Q$.)
A collinear momentum $r_c$ on the
    quark line along the $n$ direction has the scaling behavior 
\begin{eqnarray}
\label{eq:n-collinear-mom}
r_c^+ \sim Q,
\quad
r_c^- \sim Q\lambda^2,
\quad
r_{c\perp}\sim Q\lambda.
%---------------
\end{eqnarray}
%---------------

We define collinear projectors
\begin{subequations}%
%---------------
\begin{eqnarray}
%---------------
P_n &=& \frac{\slashed{n}\slashed{\bar{n}}}{4},\\
P_{\bar{n}}
&=&
\frac{\slashed{\bar{n}}\slashed{n}}{4},
%---------------
\end{eqnarray}
%---------------
where
\begin{eqnarray}
P_n+P_{\bar n}=1.
\end{eqnarray}
\end{subequations}%

The QCD Dirac field $\psi$ can be decomposed into
$n$-collinear SCET fields $\xi_n$ and $\eta_n$ by applying the
projectors $P_n$ and $P_{\bar{n}}$:
\begin{eqnarray}
\label{eq:projectors}%
\xi_n = P_n \psi,
\quad
\eta_n = P_{\bar{n}} \psi.
\end{eqnarray}
In addition, we define an $n$-collinear gluon
field $G_n^\mu$.  $\xi_n$ has homogeneous power counting in $\lambda$,
while each component of the field $G_n^\mu$ has the same power
counting as the $n$-collinear momentum in
Eq.~(\ref{eq:n-collinear-mom}).

We define an $n$-collinear covariant derivative and its left-acting
counterpart:
%---------------
\begin{subequations}
\begin{eqnarray}
%---------------
\label{eq:covariant-derivative}%
iD_n^\mu&=&i\partial^\mu +g_s G_n^\mu,\\
\label{eq:left-covariant-derivative}%
i\overleftarrow{D}_n^\mu &=& i\overleftarrow{\partial}^\mu
-g_s G_n^\mu.
%---------------
\end{eqnarray}
\end{subequations}%
%---------------
Here, $g_s = \sqrt{4\pi \alpha_s}$ is the strong coupling.
We also define the $n$-collinear Wilson line as
%---------------
\begin{eqnarray}
\label{def:collinear-Wilson}
%---------------
W_n(x) &=& P\exp\left[ig_s\int_{-\infty}^0 ds\, \bar{n}\cdot
  G_n(x+s\bar{n})\right],
%---------------
\end{eqnarray}
%---------------
which follows from our convention that the quark-gluon
vertex contributes a factor $ig_s\gamma_\mu$, where $\mu$ is the gluon
polarization index.

\section{Operator bases\label{sec:bases}}
\subsection{Building blocks from QCD \label{sec:QCD-BB}}

Let us first describe a basis of collinear operator building blocks
that follows directly from QCD.

The part of the QCD Lagrangian
\begin{eqnarray}
\mathcal{L}_\textrm{QCD}
&=& 
\bar{\psi}(i\slashed{D}-m)\psi
\end{eqnarray}
that involves the collinear SCET fields
$\xi_n$, $\eta_n$, and $G_n$, but not the soft SCET fields, is
%---------------
\begin{eqnarray}
\label{eq:QCD-lagrangian}
%---------------
\mathcal{L}_\textrm{SCET-coll}
&=&
\left(\bar\xi_n+\bar\eta_n\right)
\left(in\cdot D_n\frac{\slashed{\bar{n}}}{2}
+i\bar{n}\cdot D_n\frac{\slashed{n}}{2}
+i\slashed{D}_{n\perp}
-m\right)
\left(\xi_n+\eta_n\right),
%---------------
\end{eqnarray}
%---------------
where the covariant derivative $D_n$ is defined in
Eq.~(\ref{eq:covariant-derivative}), and we have decomposed $\psi$ as
$\psi = \xi_n + \eta_n$ by making use of the projection operators.
We can simplify this expression by making use of the projection
identities $\slashed{{n}}\xi_n = \slashed{\bar{n}}\eta_n
=\bar{\xi}_n\slashed{{n}} =\bar{\eta}_n\slashed{\bar{n}}= 0$. The
result is
%---------------
\begin{eqnarray}
\label{eq:lag-xi-eta-q}
%---------------
\mathcal{L} 
&=&
\bar\xi_n
\left(in\cdot D_n\frac{\slashed{\bar{n}}}{2}\right)
\xi_n
+
\bar\eta_n
\left(i\bar{n}\cdot D_n\frac{\slashed{{n}}}{2}\right)
\eta_n
+
\bar\xi_n
\left(i\slashed{D}_{n\perp}
-m\right)
\eta_n
+
\bar\eta_n
\left(i\slashed{D}_{n\perp}
-m\right)
\xi_n.
\phantom{XX}
%---------------
\end{eqnarray}
%---------------
Note that this Lagrangian is quadratic in the ``small component'' of the
fermion field $\eta_n$, and so we can integrate
out $\eta_n$ by making use of its equations of motion, which
are
\begin{subequations}%
\label{eq:EOM}%
%---------------
\begin{eqnarray}
%---------------
\eta_n &=& -\frac{\slashed{\bar{n}}}{2}
\frac{1}{i\bar{n}\cdot D_n}
(i\slashed{D}_{n\perp} - m) \xi_n,
\\
\bar{\eta}_n
&=&
\bar{\xi}_n
(-i\overleftarrow{\slashed{D}}_{n\perp}-m)
\frac{1}{i\bar{n}\cdot \overleftarrow{D}_n}
\frac{\slashed{\bar{n}}}{2},
%---------------
\end{eqnarray}
\end{subequations}%
where the left-acting covariant derivative
    $\overleftarrow{\slashed{D}}_{n}$ is defined in
    Eq.~(\ref{eq:left-covariant-derivative}).
%---------------
Then, inserting Eq.~(\ref{eq:EOM}) into
Eq.~(\ref{eq:lag-xi-eta-q}), we obtain
%---------------
\begin{eqnarray}
\label{eq:lagrangian-with-mass}
%---------------
\mathcal{L} 
&=&
\bar\xi_n
\left[
in\cdot D_n
+
\left(i\slashed{D}_{n\perp}
-m\right)
\frac{1}{i\bar{n}\cdot D_n}
\left(i\slashed{D}_{n\perp}
+m\right)
\right]
\frac{\slashed{\bar{n}}}{2}
\xi_n.
%---------------
\end{eqnarray}
%---------------

Now, let us consider a hard-scattering exclusive process in which one of
the external states is an $n$-collinear meson, consisting of a quark
$\cal{Q}$ and an antiquark $\cal{\bar{Q}}$ with masses $m$. The
factorization theorem for this process contains an $n$-collinear
function. Two heavy-quark propagators within the $n$-collinear function
connect it to a hard subdiagram in which the internal momenta have all
components of order $Q$. In full QCD, the Dirac operators $\psi$ and
$\bar\psi$ terminate these propagators. When one decouples
longitudinally polarized $n$-collinear gluons from mesons with
non-$n$-collinear momenta and from the hard-scattering subdiagram, using
standard methods \cite{Collins:1989gx}, Wilson-line factors appear.
Since $\psi$ does not have definite power counting in $\lambda$, we do
not take $W_n^\dagger\psi$ and $\bar\psi W_n$ as building
blocks. Instead, we decompose $\psi$ into the SCET fields $\xi_n$ and
$\eta_n$ [Eq.~(\ref{eq:projectors})], which have definite power counting
in $\lambda$, and arrive at the building blocks
\begin{subequations}%
\label{eq:building-blocks-1}%
\begin{eqnarray}
\label{eq:buliding-blocks-1-chi}
\chi_n&=&W_n^\dagger \xi_n,\\
\bar\chi_n&=&\bar\xi_nW_n,\\
\phi_n&=&W_n^\dagger \eta_n = -W_n^\dagger
\frac{\slashed{\bar{n}}}{2}
\frac{1}{i\bar{n}\cdot D_n}
(i\slashed{D}_{n\perp} - m) \xi_n,\\
\bar\phi_n&=&\bar\eta_n W_n = \bar{\xi}_n
(-i\overleftarrow{\slashed{D}}_{n\perp}-m)
\frac{1}{i\bar{n}\cdot \overleftarrow{D}_n}
\frac{\slashed{\bar{n}}}{2}W_n,
\label{eq:building-blocks-1d}%
\end{eqnarray}
\end{subequations}
where we have used the equations of motion in Eq.~(\ref{eq:EOM}) in the
last two lines.

The last two lines of Eq.~(\ref{eq:building-blocks-1}) are of order
$\lambda$ relative to the first two lines.  In order to complete the set
of building blocks in relative order $\lambda$, we need to consider
processes in which, in addition to the $\cal{Q}$ and $\cal{\bar Q}$
propagators, an $n$-collinear-transverse-gluon propagator connects the
$n$-collinear function to the hard subdiagram. The gluon propagator with
transverse polarization $\nu$ is terminated by the operator
$(1/g_s)iD_{n\perp\nu}$ \cite{Anikin:2009bf}. When one decouples
longitudinally polarized $n$-collinear gluons from mesons with
non-$n$-collinear momenta and from the hard subdiagram, using
standard methods, Wilson lines appear, and we arrive at the building
blocks \cite{Marcantonini:2008qn, Beneke:2017ztn}
\begin{subequations}%
\label{eq:building-blocks-2}%
\begin{eqnarray}
{\cal G}_{n\perp\nu}&=&
[W_n^\dagger iD_{n\perp\nu} W_n],\\
{\cal G}_{n\perp\nu}^\dagger&=&
[W_n^\dagger
(- i\overleftarrow{D}_{n\perp\nu}) W_n],
\end{eqnarray}
\end{subequations}%
where the derivatives act only within the square brackets. These
building blocks can be used with the building blocks $\chi_n$ and
$\bar\chi_n$ in the first two lines of Eq.~(\ref{eq:building-blocks-1})
to construct collinear functions of relative order $\lambda$.  We call
these building blocks the ``direct-QCD building-block basis.''

\subsection{Modified building blocks \label{sec:modified-bb}}

Now let us describe a modification of the operator building blocks.
Using the Wilson-line identities \cite{Beneke:2002ph}
\begin{subequations}%
\label{eq:Wilson-line-identity}
\begin{eqnarray}%
W_n^\dagger \frac{1}{i\bar n\cdot D_n}&=&\frac{1}{i\bar{n}\cdot
  \partial}
W_n^\dagger,\\
\frac{1}{i\bar n\cdot D_n}W_n&=&W_n\frac{1}{i\bar{n}\cdot \partial},
\end{eqnarray}%
\end{subequations}%
we can re-write the building blocks $\phi_n$ and $\bar\phi_n$
[Eq.~(\ref{eq:building-blocks-1})] as follows:
%---------------
\begin{subequations}%
\label{eq:exp-phin}
\begin{eqnarray}
%---------------
\label{eq:exp-phina}%
\phi_n
&=&
-\frac{\slashed{\bar{n}}}{2}
W_n^\dagger
\frac{1}{i\bar{n}\cdot D_n}
(i\slashed{D}_{n\perp} - m) \xi_n
\nonumber \\
&=&
-\frac{\slashed{\bar{n}}}{2}
\frac{1}{i\bar{n}\cdot \partial}
W_n^\dagger i\slashed{D}_{n\perp} \xi_n
+m\frac{\slashed{\bar{n}}}{2}
\frac{1}{i\bar{n}\cdot \partial}
W_n^\dagger\xi_n\nonumber\\
&=&-\frac{\slashed{\bar{n}}}{2} \frac{1}{i\bar n\cdot \partial}[W_n^\dagger
    i\slashed{D}_{n\perp} W_n] \chi_n
-\frac{\slashed{\bar{n}}}{2}\frac{1}{i\bar n \cdot\partial}
\left(i\slashed{\partial}_\perp-m\right) \chi_n
\nonumber \\
&=&
-\frac{\slashed{\bar{n}}}{2} \frac{1}{i\bar n\cdot \partial}
\slashed{\cal G}_{n\perp} \chi_n
-\frac{\slashed{\bar{n}}}{2}\frac{1}{i\bar n \cdot\partial}
\left(i\slashed{\partial}_\perp-m\right) \chi_n,
\\
\bar\phi_n
&=&
\bar{\xi}_n
(-i\overleftarrow{\slashed{D}}_{n\perp}-m)
\frac{1}{i\bar{n}\cdot \overleftarrow{D}_n}
W_n
\frac{\slashed{\bar{n}}}{2}
\nonumber \\
&=&
\bar{\xi}_n
(-i\overleftarrow{\slashed{D}}_{n\perp})
W_n
\frac{1}{i\bar{n}\cdot \overleftarrow{\partial}}
\frac{\slashed{\bar{n}}}{2}
-m \bar{\xi}_n W_n
\frac{1}{i\bar{n}\cdot  \overleftarrow{\partial}}
\frac{\slashed{\bar{n}}}{2}\nonumber\\
&=&\bar\chi_n[W_n^\dagger (-i\overleftarrow{\slashed{D}}_{n\perp}) W_n] 
\frac{1}{i \bar{n}\cdot \overleftarrow{\partial} } \frac{\slashed{\bar n}}{2}
+\bar\chi_n \left(-i\overleftarrow{\slashed{\partial}}_\perp -m\right)
\frac{1}{i\bar n\cdot\overleftarrow{\partial}} \frac{\slashed{\bar n}}{2}
\nonumber\\
&=&\bar\chi_n 
\slashed{\cal G}_{n\perp}^\dagger
\frac{1}{i \bar{n}\cdot \overleftarrow{\partial} } \frac{\slashed{\bar n}}{2}
+\bar\chi_n \left(-i\overleftarrow{\slashed{\partial}}_\perp -m\right)
\frac{1}{i\bar n\cdot\overleftarrow{\partial}} \frac{\slashed{\bar n}}{2},
\label{eq:exp-phinb}
%---------------
\end{eqnarray}
\end{subequations}%
%---------------
where ${\cal G}_{n\perp}$ and ${\cal G}_{n\perp}^\dagger$ are given in
    Eq.~(\ref{eq:building-blocks-2}).

We see that the effect of the identities in
Eq.~(\ref{eq:Wilson-line-identity}) is to remove the interaction with
the gluon field from the covariant derivative in the denominator and to
cause the remaining ordinary derivative to act on the Wilson line, as
well as on other factors in the expressions for $\phi_n$ and
$\bar\phi_n$. In effect, this moves the interaction from the covariant
derivative in the denominator to the Wilson line.  As we will see, this
complicates the identification of the quark self-energy contributions.
The factors $(\slashed{\bar{n}}/2)[1/(i\bar n \cdot \partial)]$ and
$[1/(i\bar n \cdot \overleftarrow{\partial})](\slashed{\bar{n}}/2)$
involve only a large (plus) momentum component, and so they can be
absorbed into the hard factor (Wilson coefficient). We will return to
these points when we discuss calculations of the collinear functions for
the modified building blocks in Sec.~\ref{sec:calc-mod-blocks}.  As we
will see, this re-arrangement of the building-block basis leads to
difficulties in the treatment of the external self-energy diagrams.

Now, we can consider the building blocks to be 
\begin{equation}
\label{eq:BB-modified}%
\chi_n,\;
\bar\chi_n,\;
{\cal G}_{n\perp\nu},\;
{\cal G}_{n\perp\nu}^\dagger,\;
\end{equation}
supplemented by the factors $\partial_\perp$,
$\overleftarrow{\partial}_\perp$, and $m$, which introduce factors of
$\lambda$.  We call these building blocks the ``modified
    building-block basis.''  This basis of building blocks was deduced
in Refs.~\cite{Feige:2017zci,Beneke:2017ztn,Beneke:2018rbh} from the
requirement of gauge covariance, without direct reference to QCD.

\section{Calculation of the quark self-energy\label{sec:self-energy-calcs}}

In this section, we illustrate the difficulty that can arise in SCET in
  identifying quark self-energies on external legs. We compare
  calculations of the quark self-energy on an external leg in full
  QCD, in the SCET building-block basis that follows directly from QCD
  in Eqs.~(\ref{eq:building-blocks-1}) and (\ref{eq:building-blocks-2})
  and in the modified SCET building-block basis in
  Eq.~(\ref{eq:BB-modified}).

For definiteness, we carry out the calculations in the Feynman
gauge. However, we expect the difficulties that we illustrate to occur
more generally. Of course, quark self energies are not gauge
invariant. Neither are Green's functions with amputated external
complete quark propagators. Gauge invariance is achieved when one forms
the $S$-matrix by multiplying the amputated Green's functions by the
square roots of wave-function renormalizations (from the quark
self-energies).\footnote{See, for example, pp.~344--346 of
Ref.~\cite{Sterman:1993hfp}.}

\subsection{QCD calculation}
%---------------------------------------
\begin{figure}
\centering
\includegraphics[scale=0.6]{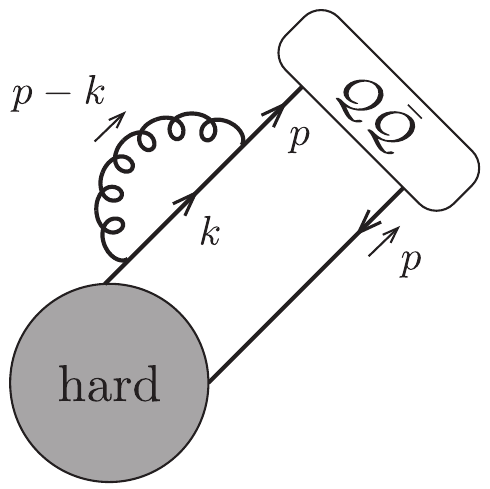}
\caption{\label{fig:QCDself} A diagram for a self-energy correction on
  an outgoing external quark-antiquark (${\cal Q} \bar {\cal Q}$)
  meson. The solid lines represent a quark or an antiquark, the wavy line
      represents a gluon, the rectangular region that is labeled ``${\cal
        Q}\bar{\cal Q}$'' represents the external ${\cal Q}\bar{\cal Q}$
      state, and the circle labeled ``hard'' represents the hard
      subdiagram.}
\end{figure}
%---------------------------------------
In Fig.~\ref{fig:QCDself}, we show a diagram for a self-energy
correction on an outgoing external quark-antiquark (${\cal Q} \bar {\cal
  Q}$) meson. For simplicity, we consider only self-energy corrections
on the quark line. The circle labeled ``hard'' represents the hard
subdiagram in a hard-scattering process, for example,
virtual-photon-meson scattering.  It is a matrix in the Dirac and color
indices of the external ${\cal Q}$ and $\bar{\cal Q}$.  We have
suppressed external legs on the hard subdiagram that correspond to
additional mesons or to off-shell particles that carry a hard
momentum. Recall that all of the internal propagators in the hard
subdiagram carry hard momenta, all of whose components scale as
$\lambda^0 Q$.

The expression in QCD for the quark self-energy on an outgoing external
$\cal{Q}$ line that corresponds to the diagram in Fig.~\ref{fig:QCDself}
is given by
%---------------
\begin{eqnarray}
\label{eq:QCD-self}%
%---------------
i\mathcal{A}_\textrm{QCD}
&=&
-i
g_s^2C_F
\textrm{Tr}
\int_k
\frac{
v(p)
\bar{u}(p)
\gamma^\rho
(\slashed{k}+m)
\gamma_\rho
(\slashed{p}+m)
H_\textrm{QCD}
}
{
(k^2-m^2+i\varepsilon)
[(p-k)^2+i\varepsilon]
(p^2-m^2+i\varepsilon)
},
%---------------
\end{eqnarray}
%---------------
where $\bar{u}(p)$ and $v(p)$ are the spinors for the external quark and
antiquark states $\cal{Q}$ and ${\cal \bar{Q}}$, respectively, and
$H_\textrm{QCD}$ is the Dirac- and color-index-valued matrix that
corresponds to the hard subdiagram in Fig.~\ref{fig:QCDself}.  We
suppress the color and Dirac indices on $H_\textrm{QCD}$ and the
dependence of $H_\textrm{QCD}$ on external momenta.  The trace is over
the Dirac and color indices, and $p=(p^+,p^-,p_\perp)$ is the
momentum of the outgoing quark. We hold $p^-$ slightly off the mass
shell in order to avoid the singularity in the denominator of
Eq.~(\ref{eq:QCD-self}) at $p^2=m^2$.  We use the notation
%---------------
\begin{eqnarray}
%---------------
\int_k \equiv \mu^{2\epsilon}\int \frac{d^D k}{(2\pi)^D},
%---------------
\end{eqnarray}
%---------------
where $D$ is the dimensionality of space-time in dimensional regularization.
The expression for $i{\cal A}_{\rm QCD}$ can be expanded in powers of
    $\lambda$ with the result
\begin{eqnarray}
\label{eq:QCD-power-expanded}
i\mathcal{A}_\textrm{QCD}
&=&
-
ig_s^2C_F
\textrm{Tr}
\int_k
\frac{
P_n
v(p)
\bar{u}(p)
\left(
N+N_\perp
\right)
H_\textrm{QCD}
}
{
(k^2-m^2+i\varepsilon)
[(p-k)^2+i\varepsilon]
(p^2-m^2+i\varepsilon)},
\end{eqnarray}
where
\begin{subequations}
\begin{eqnarray}
\label{eq:def-of-numerator}
N
&=&
Dm^2
+
m[Dp^+ - (D-2)k^+]\frac{\slashed{n}}{2}
-
(D-2) k^- p^+ \frac{\slashed{\bar{n}}\slashed{n}}{4}
\nonumber \\
&&
+
m[Dp^- - (D-2)k^-]\frac{\slashed{\bar{n}}}{2}
-
(D-2) k^+ p^- \frac{\slashed{n}\slashed{\bar{n}}}{4},
\end{eqnarray}
and
\begin{eqnarray}
\label{eq:def-of-numerator-perp}
N_\perp
&=&
Dm\slashed{p}_\perp -
(D-2)
\left(
m\slashed{k}_\perp
+
\slashed{k}_\perp \slashed{p}_\perp
\right)
+
(D-2)
\left(
k^+ \slashed{p}_\perp
-
p^+\slashed{k}_\perp 
\right)
\frac{\slashed{{n}}}{2}
\nonumber \\
&&
+
(D-2)
\left(
k^- \slashed{p}_\perp
-
p^- \slashed{k}_\perp
\right)
\frac{\slashed{\bar{n}}}{2}.
\end{eqnarray}
\end{subequations}
Here we have introduced the projector $P_n$ in
    order to compare with SCET results (below).  In
Eq.~\eqref{eq:QCD-power-expanded}, the denominators scale as
$\lambda^{-6}$, while the integration volume contributes
$\lambda^4$. The spinor factor, $v(p)\bar{u}(p)$, contributes with
relative power $\lambda^0$ when contracted with $\slashed{\bar{n}}$ on
the right (as in $\xi_n \bar\xi_n$ structures) and with relative power
$\lambda$ when contracted with $\slashed{n}$ to the right (as in $\xi_n
\bar\eta_n$ structures). If we omit powers of $\lambda$ from the hard
factor $H_\textrm{QCD}$, then we find that the first, second and third
terms of Eq.~\eqref{eq:def-of-numerator} contribute at power
$\lambda^0$, while the fourth and fifth terms contribute at power
$\lambda^1$.  We also find that the first, second, third terms of
Eq.~\eqref{eq:def-of-numerator-perp} contribute at power $\lambda^0$,
while the fourth term contributes at power $\lambda^1$.

\subsection{Calculation using the direct-QCD building blocks}
\label{subsec:direct-QCD}

Using the building blocks that follow directly from QCD, we can
construct a collinear function at leading order in $\lambda$:
\begin{equation}
\label{eq:collinear-fn-E}%
E(u_1,u_2)
\equiv
\frac{(2p^+)^2}{(2\pi)^2}
\int ds_1ds_2\,
e^{-i(s_1u_1+s_2u_2)(2p^+)}
\langle {\cal Q}{\cal \bar{Q}}|
T\,
\chi_n(s_2 \bar{n})
\bar{\chi}_n(s_1\bar{n})
|0\rangle,
\end{equation}
where we take the state $\langle {\cal Q}{\cal \bar Q}|$ to consist of
    a quark and an antiquark, each with momentum $p$, as in the QCD
    calculation, and we have suppressed the color and spin indices on
    the quark fields.

We can also construct a collinear function at the first subleading order
in $\lambda$:
\begin{equation}
\label{eq:collinear-fn-J}%
J(u_1,u_2)
\equiv
\frac{(2p^+)^2}{(2\pi)^2}
\int ds_1ds_2\,
e^{-i(s_1u_1+s_2u_2)(2p^+)}
\langle {\cal Q}{\cal \bar Q}|
T\,
\chi_n(s_2 \bar{n})
\bar{\phi}_n(s_1\bar{n})
|0\rangle,
\end{equation}
where $\bar{\phi}_n$ is given in Eq.~\eqref{eq:building-blocks-1} [as
    opposed to Eq.~\eqref{eq:exp-phin}].  We do not show the
  charge-conjugate collinear function involving $\bar\chi_n$ and
  $\phi_n$ or a collinear function involving ${\cal G}_{n\perp\nu}$
  since we do not need them for the calculation of the self energy on
  the external ${\cal Q}$ leg.

%---------------------------------------
\begin{figure}
\centering
\includegraphics[scale=0.6]{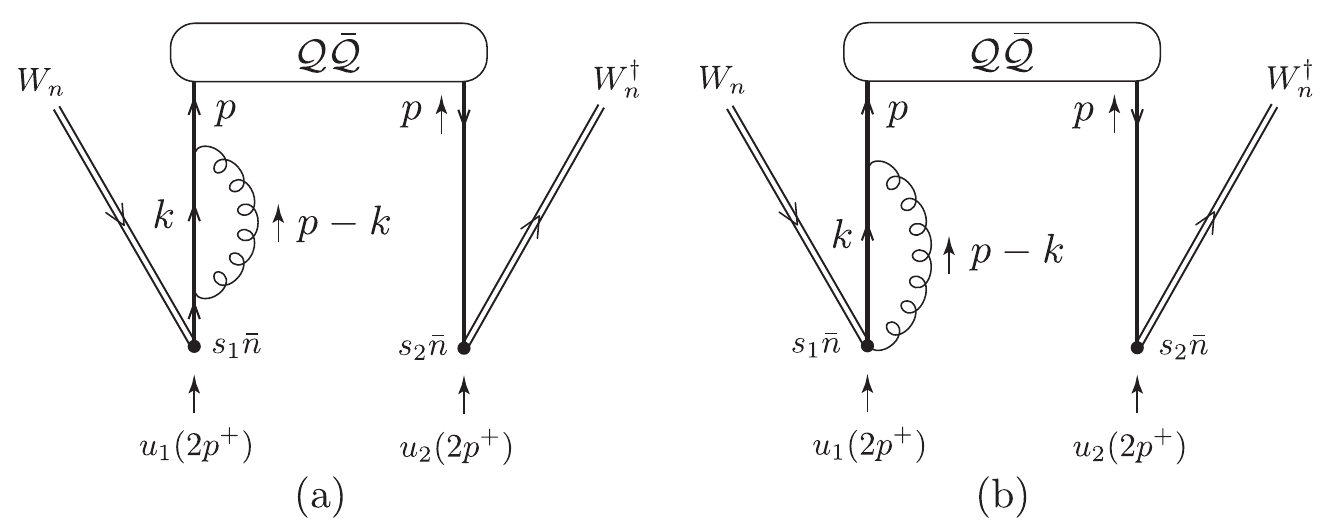}
\caption{\label{fig:SCETself} Self-energy Feynman diagrams that could
  contribute to collinear functions. Note that, in this figure, and
  throughout this paper, we display Wilson lines explicitly, instead of
  incorporating them into definitions of operator vertices.  }
\end{figure}
%---------------------------------------
We now compute the self-energy contribution at leading order in
$\lambda$ from the collinear function $E(u_1,u_2)$.  Note that
$E(u_1,u_2)$ contains some subleading contributions, as well, that arise
from subleading powers in the full-QCD or SCET interaction Lagrangians.
It is simplest to carry out this calculation by using the full-QCD
Feynman rules, supplemented by projectors that relate the fields
$\chi_n$ and $\bar\chi_n$ to $\psi$ and $\bar\psi$
[Eq.~(\ref{eq:projectors})].  There are the two self-energy diagrams on
the ${\cal Q}$ line, which are shown in Fig.~\ref{fig:SCETself}.  In
order to clarify the diagrammatic topologies, in this figure, and
throughout this paper, we explicitly display the collinear Wilson lines
that are associated with the collinear quark building blocks $\chi_n$
and $\bar{\chi}_n$, which are defined in
Eq.~\eqref{eq:building-blocks-1}.  This is in contrast with the practice
elsewhere in the SCET literature in which Wilson-line contributions are
contained diagrammatically within the vertices.  Note that gluons that
attach to Wilson lines do not correspond to external self-energy
diagrams in QCD because the Wilson-line gluon attachments were
originally attachments to non-$n$-collinear lines or to hard lines in
QCD.  Therefore, we do not consider contributions in which gluons attach
to Wilson lines.  Note that diagram (b) in Fig.~\ref{fig:SCETself} does
not contribute to $E(u_1,u_2)$ because the only gluon fields that are
associated with the point $s_1\bar{n}$ are those in the Wilson line.
Hence,
%---------------
\begin{equation}
%---------------
E(u_1,u_2)\big|_{(\textrm{b})}=0,
%---------------
\end{equation}
%---------------
where the subscript (b) on $E(u_1,u_2)$ refers to diagram (b) in
    Fig.~\ref{fig:SCETself}.  The result for diagram (a) in
Fig.~\ref{fig:SCETself} is
%---------------
\begin{eqnarray}
%---------------
E(u_1,u_2)\big|_{\textrm{(a)}}
&=&
-
ig_s^2C_F\delta(u_1-\tfrac{1}{2}) \delta(u_2-\tfrac{1}{2})
\int_k
\frac{
P_n 
v(p)\bar{u}(p)
\gamma^\rho
(\slashed{k}+m)
\gamma_\rho
(\slashed{p}+m)
P_{\bar{n}}
}
{(k^2-m^2+i\varepsilon)
[(p-k)^2+i\varepsilon]
(p^2-m^2+i\varepsilon)},
\nonumber \\
%---------------
\end{eqnarray}
%---------------
where we have suppressed the Dirac and color indices on
$E(u_1,u_2)$. Simplifying the numerator, we obtain
%---------------
\begin{eqnarray}
\label{eq:Eu1u2_a}
%---------------
E(u_1,u_2)\big|_{\textrm{(a)}}
&=&
-
ig_s^2C_F\delta(u_1-\tfrac{1}{2}) \delta(u_2-\tfrac{1}{2})
\int_k
\frac{
P_n 
v(p)\bar{u}(p)
\left(
N^{E_{\textrm{(a)}}}
+
N^{E_{\textrm{(a)}}}_\perp
\right)
}
{(k^2-m^2+i\varepsilon)
[(p-k)^2+i\varepsilon]
(p^2-m^2+i\varepsilon)},
\nonumber \\
%---------------
\end{eqnarray}
%---------------
where
\begin{eqnarray}
N^{E_{\textrm{(a)}}}
&=&
m[Dp^+-(D-2)k^+]\frac{\slashed{n}}{2}
-
[(D-2)p^+k^--Dm^2]\frac{\slashed{\bar{n}}\slashed{n}}{4},
\nonumber \\
N^{E_{\textrm{(a)}}}_\perp
&=&
(D-2)
\left(
k^+ \slashed{p}_\perp
-
p^+ \slashed{k}_\perp
\right)
\frac{\slashed{n}}{2}
+
\left[
Dm\slashed{p}_\perp -(D-2)m \slashed{k}_\perp
-(D-2)\slashed{k}_\perp \slashed{p}_\perp
\right]
\frac{\slashed{\bar{n}}\slashed{n}}{4}.
\end{eqnarray}

Similarly, we can compute the self-energy contribution at the subleading
order in $\lambda$ from the collinear function $J(u_1,u_2)$. Again, we
make use of the QCD Feynman rules, supplemented by projectors that
relate the fields $\chi_n$ and $\bar\phi_n$ to $\psi$ and $\bar\psi$. We
also make use of the Feynman rules from the definition of $\bar\phi_n$,
which is given in Eq.~(\ref{eq:building-blocks-1}) [as opposed to
  Eq.~(\ref{eq:exp-phin})].  The result for diagram (a) in
Fig.~\ref{fig:SCETself} is
%---------------
\begin{eqnarray}
%---------------
J(u_1,u_2)\big|_{\textrm{(a)}}
&=&
-
ig_s^2C_F
\frac{(2p^+)^2}{(2\pi)^2}
\int ds_1ds_2\,
e^{-i(s_1u_1+s_2u_2)(2p^+)}
e^{ip^+s_1}
e^{ip^+s_2}
\nonumber \\
&&
\times
\int_k
\frac{
P_n 
v(p)\bar{u}(p)
\gamma^\rho
(\slashed{k}+m)
\gamma_\rho
(\slashed{p}+m)
P_{\bar{n}}
\left[
(\slashed{p}_\perp-m)
\frac{1}{-p^+}
\right]
\frac{\slashed{\bar{n}}}{2}
}
{(k^2-m^2+i\varepsilon)
[(p-k)^2+i\varepsilon]
(p^2-m^2+i\varepsilon)}.
%---------------
\end{eqnarray}
%---------------
The factors in the square brackets arise from the field $\bar{\phi}_n$ that
    appears in the definition of $J(u_1,u_2)$.  Simplifying the
numerator, we obtain
%---------------
\begin{eqnarray}
%---------------
J(u_1,u_2)\big|_{\textrm{(a)}}
&=&
-
ig_s^2C_F\delta(u_1-\tfrac{1}{2})\delta(u_2-\tfrac{1}{2})
\nonumber\\
&&
\times
\int_k
\frac{
P_n 
v(p)\bar{u}(p)
\frac{1}{p^+}
\left(
N^{J_{\textrm{(a)}}}
+
N^{J_{\textrm{(a)}}}_\perp
\right)
}
{(k^2-m^2+i\varepsilon)
[(p-k)^2+i\varepsilon]
(p^2-m^2+i\varepsilon)},
%---------------
\end{eqnarray}
%---------------
where
\begin{subequations}%
\begin{eqnarray}%
N^{J_{\textrm{(a)}}}
&=&
m^2[Dp^+-(D-2)k^+]\frac{\slashed{n}\slashed{\bar{n}}}{4}
+
m
[Dm^2-(D-2)p^+k^-]\frac{\slashed{\bar{n}}}{2},
\\
N_\perp^{J_{\textrm{(a)}}}
&=&
\left\{
p^+\left[
Dm\slashed{p}_\perp
-
(D-2)m\slashed{k}_\perp
-
(D-2)
\slashed{k}_\perp \slashed{p}_\perp
\right]
+
(D-2)k^+ {p}_\perp^2
\right\}
\frac{\slashed{n}\slashed{\bar{n}}}{4}
\nonumber \\
&&
+
\left[
(D-2)p^+ k^- \slashed{p}_\perp
-
Dm{p}_\perp^2 
-
(D-2)(m^2-{p}_\perp^2)\slashed{k}_\perp
\right]
\frac{\slashed{\bar{n}}}{2}.
\end{eqnarray}%
\end{subequations}%
Again, gluons that attach to the Wilson lines
    do not contribute to the $\cal{Q}$ self energy.  In this work, we
define $p_\perp^2 = -\bm{p}_\perp^2$, where $p_\perp$ denotes the
Minkowski four-vector and $\bm{p}_\perp$ its corresponding Euclidean
vector. We adopt an analogous convention for $k_\perp$.

The contribution of diagram (b) in Fig.~\ref{fig:SCETself}
to the collinear function $J(u_1,u_2)$ is given by
%---------------
\begin{eqnarray}
%---------------
J(u_1,u_2)\big|_{\textrm{(b)}}
&=&
ig_s^2C_F
\frac{(2p^+)^2}{(2\pi)^2}
\int ds_1ds_2\,
e^{-i(s_1u_1+s_2u_2)(2p^+)}
e^{ip^+s_1}
e^{ip^+s_2}
\nonumber \\
&&
\times
\int_k
\frac{
P_n 
v(p)\bar{u}(p)
\gamma^\rho
(\slashed{k}+m)
P_{\bar{n}}
\left[
(\gamma_{\perp\rho})
\frac{1}{-p^+}
+
(\slashed{k}_\perp-m)
\left(
\frac{1}{-k^+}
\bar{n}_\rho
\frac{1}{-p^+}
\right)
\right]
\frac{\slashed{\bar{n}}}{2}
}
{(k^2-m^2+i\varepsilon)
[(p-k)^2+i\varepsilon]}.
\nonumber \\
%---------------
\end{eqnarray}
%---------------
In deriving this expression, we have used the following identity
for the inverse covariant derivative in $\bar{\phi}_n$ in order to obtain
    the order $g_s$ contribution of the gluon field at the vertex at
$s_1\bar{n}$:
%---------------
\begin{eqnarray}
%---------------
\frac{1}{i\bar n\cdot \overleftarrow{D}_n}
=
\frac{1}{i\bar n\cdot \overleftarrow{\partial} - g_s \bar{n}\cdot G_n}
=
\frac{1}{i\bar n\cdot \overleftarrow{\partial}}
+
\frac{1}{i\bar n\cdot \overleftarrow{\partial}}
g_s \bar{n}\cdot G_n
\frac{1}{i\bar n\cdot \overleftarrow{\partial}}
+
O(g_s^2).
%---------------
\end{eqnarray}
%---------------
Simplifying the numerator, we obtain
%---------------
\begin{eqnarray}
\label{eq:JB-in-Feynman}
%---------------
J(u_1,u_2)\big|_{\textrm{(b)}}
&=&
ig_s^2C_F
\delta(u_1-\tfrac{1}{2})\delta(u_2-\tfrac{1}{2})
\int_k
\frac{
P_n v(p)\bar{u}(p)
\frac{1}{p^+}
\left(
N^{J_{\textrm{(b)}}}
+
N^{J_{\textrm{(b)}}}_\perp
\right)
}
{(k^2-m^2+i\varepsilon)
[(p-k)^2+i\varepsilon]},
%---------------
\end{eqnarray}
%---------------
where
\begin{subequations}%
\begin{eqnarray}%
N^{J_{\textrm{(b)}}}
&=&
(D-2)k^+
\frac{\slashed{n}\slashed{\bar{n}}}{4}
-D m \frac{\slashed{\bar{n}}}{2},
\\
N_\perp^{J_{\textrm{(b)}}}
&=&
(D-2)\slashed{k}_\perp \frac{\slashed{\bar{n}}}{2}.
\end{eqnarray}
\end{subequations}%

We define the full amplitudes, including the Wilson coefficients
    (matching coefficients), for each collinear function as
\begin{eqnarray}
\label{eq:Wilson-EJ}
i\mathcal{A}_E &\equiv&\textrm{Tr}\int du_1du_2
\left[E(u_1,u_2)|_{\textrm{(a)}}\right]H_\textrm{QCD},
\nonumber \\
i\mathcal{A}_{J_{\textrm{(a,b)}}} &\equiv&\textrm{Tr}\int du_1du_2
\left[J(u_1,u_2)|_{\textrm{(a,b)}}\right]H_\textrm{QCD},
\end{eqnarray}
where the trace is over Dirac and color indices.  The factor
$H_\textrm{QCD}$ is the hard factor that appears in the full-QCD
amplitudes. In the case of the collinear functions $E(u_1,u_2)$ and
$J(u_1,u_2)$, it is the Wilson coefficient for those collinear
functions, which is determined by matching the SCET amplitude to the
full-QCD amplitude in Eq.~(\ref{eq:QCD-self}).  We suppress the color
and Dirac indices on $H_\textrm{QCD}$ and also suppress the dependence
of $H_\textrm{QCD}$ on the external momenta, including the momenta
$u_1(2p^+)$ and $u_2(2p^+)$ that are associated with the outgoing ${\cal
  Q}$ and $\bar{\cal Q}$.  The sum of the three nonzero contributions to
the SCET amplitude for the self-energy correction is 
%---------------
\begin{eqnarray}
\label{eq:sum-of-E-J}
%---------------
i\mathcal{A}_E + i\mathcal{A}_{J_{\textrm{(a)}}} + i\mathcal{A}_{J_{\textrm{(b)}}}
&=&
-
ig_s^2C_F
\textrm{Tr}
\int_k
\frac{
P_n 
v(p)\bar{u}(p)
\left(
N^{E+J}
+
N_\perp^{E+J}
\right)
H_\textrm{QCD}
}
{(k^2-m^2+i\varepsilon)
(p^2-m^2+i\varepsilon)
[(p-k)^2+i\varepsilon]},
%---------------
\end{eqnarray}
%--------------- 
where
\begin{eqnarray}
N^{E+J} &=&
Dm^2
-
(D-2) k^+ p^- \frac{\slashed{n}\slashed{\bar{n}}}{4}
-
(D-2) k^- p^+ \frac{\slashed{\bar{n}}\slashed{n}}{4}
\nonumber \\
&&
+
m[Dp^- - (D-2)k^-]\frac{\slashed{\bar{n}}}{2}
+
m[Dp^+ - (D-2)k^+]\frac{\slashed{n}}{2},
\nonumber \\
N^{E+J}_\perp &=&
Dm\slashed{p}_\perp -
(D-2)
\left(
m\slashed{k}_\perp
+
\slashed{k}_\perp \slashed{p}_\perp
\right)
\nonumber \\
&&
+
(D-2)
\left(
k^+ \slashed{p}_\perp
-
p^+\slashed{k}_\perp 
\right)
\frac{\slashed{{n}}}{2}
+
(D-2)
\left(
k^- \slashed{p}_\perp
-
p^- \slashed{k}_\perp
\right)
\frac{\slashed{\bar{n}}}{2}.
\end{eqnarray}
Comparing Eq.~\eqref{eq:QCD-power-expanded} and Eq.~\eqref{eq:sum-of-E-J},
we find that the complete amplitude, including the factor from the hard
subdiagram, yields
%---------------
\begin{eqnarray}
\label{eq:QCD-SCET-compare}%
%---------------
i\mathcal{A}_E + i\mathcal{A}_{J_{\textrm{(a)}}} + i\mathcal{A}_{J_{\textrm{(b)}}}=
i\mathcal{A}_\textrm{QCD}.
%---------------
\end{eqnarray}
%---------------

We conclude that the contributions to the collinear functions $E(u_1,u_2)$ and
$J(u_1,u_2)$ from the diagrams in Fig.~\ref{fig:SCETself} account fully for
the self-energy diagrams on the external ${\cal Q}$ leg.  

\subsection{Calculation using the modified building blocks 
\label{sec:calc-mod-blocks}}

\subsubsection{Collinear functions}

Now let us calculate the self-energy contributions in SCET by making use
of collinear functions that follow from the modified operator building
blocks in Eq.~(\ref{eq:BB-modified}).  Since $\chi_n$ and $\bar{\chi}_n$
are among these building blocks, we can again construct a collinear
function $E(u_1,u_2)$ [Eq.~(\ref{eq:collinear-fn-E})].  However, we no
longer have the building blocks $\phi_n$ and $\bar{\phi}_n$, and so we
cannot construct the collinear function $J(u_1,u_2)$
[Eq.~(\ref{eq:collinear-fn-J})].  Instead, we use the building blocks
$\chi_n$, $\bar\chi_n$, and $\mathcal{G}_{n\perp\nu}$ to construct
collinear functions at subleading power (order $\lambda$). The collinear
functions that are relevant for the self-energy calculation follow
directly from the expressions in Eq.~(\ref{eq:exp-phin}). They are
\begin{eqnarray}
\label{eq:F_and_G}
F(u_1, u_2)
&=&
\frac{(2p^+)^2 }{(2\pi)^2}
\int ds_1ds_2\,
e^{-i(s_1u_1+s_2u_2)(2p^+)}
\langle {\cal Q} \bar{\cal Q}| T\,
\chi_n(s_2 \bar{n})
\left[
(-i\slashed\partial_{\perp}-m)
\bar{\chi}_n(s_1\bar{n})
\right]
|0\rangle,
\nonumber \\
G(u_1,u_2)
&=&
\frac{(2p^+)^3 }{(2\pi)^3}
\int ds_1ds_2\,
e^{-i(s_1u_1+s_2u_2)(2p^+)}
\langle {\cal Q} \bar{\cal Q}| T\,
\chi_n(s_2 \bar{n})
[\bar{\chi}_n
\slashed{\mathcal{G}}_{n\perp}^\dagger
](s_1 \bar{n})
|0\rangle.
\end{eqnarray}
Here, and throughout the remainder of this paper, we take the external
    states of the collinear functions to be full-QCD quark states.  Of
    course, we could expand these states in terms of SCET collinear
    states.  However, it is simpler, in comparing with the full-QCD
    results for the self-energy contributions, to retain the
    full-QCD states.

Let us initially consider the contributions to these collinear functions
that have the explicit topology of a self-energy diagram.

\subsubsection{Calculation of the diagrams that have a self-energy topology}

The contribution to $F(u_1,u_2)$ from Fig.~\ref{fig:SCETself}(a) can be
found from the expression for $E(u_1,u_2)$ by applying
$(-i\slashed\partial_{\perp}-m)$ to the external quark field
$\bar{\chi}_n$ at $s_1\bar{n}$.  The result is
\begin{eqnarray}
F(u_1, u_2)\big|_{\textrm{(a)}}
&=&
E(u_1,u_2)\big|_{\textrm{(a)}}
(\slashed{p}_\perp-m)
\nonumber \\
&=&
-
ig_s^2C_F\delta(u_1-\tfrac{1}{2}) \delta(u_2-\tfrac{1}{2})
\int_k
\frac{
P_n 
v(p)\bar{u}(p)
\gamma^\rho
(\slashed{k}+m)
\gamma_\rho
(\slashed{p}+m)
P_{\bar{n}}
(\slashed{p}_\perp-m)
}
{(k^2-m^2+i\varepsilon)
[(p-k)^2+i\varepsilon]
(p^2-m^2+i\varepsilon)}.
\nonumber \\
\end{eqnarray}
Note that $F$ receives no contribution from Fig.~\ref{fig:SCETself}(b)
because the only gluon fields that are associated with the point
$s_1\bar{n}$ are those in the Wilson line.  The Wilson coefficient for
the collinear function $F(u_1, u_2)$ follows from the expressions in
Eq.~\eqref{eq:exp-phin}. As we have mentioned after
Eq.~\eqref{eq:exp-phin}, these expressions imply that there is a factor
$(\slashed{\bar{n}}/2)[1/(i\bar n \cdot \partial)]$ that should be
associated with the Wilson coefficient, in addition to the factor from
the hard subdiagram $H_\textrm{QCD}$. Hence, we find that the Wilson
coefficient for the collinear function $F$ is
\begin{equation}
\label{eq:hard_F}
H_F = \frac{1}{-p^+}\frac{\slashed{\bar{n}}}{2}H_\textrm{QCD}.
\end{equation}
Then, we obtain the following self-energy amplitude for $F(u_1, u_2)$:
\begin{eqnarray}
\label{eq:amp_from_F}
i\mathcal{A}_F
&=&
{\rm Tr}\int du_1 du_2 
F(u_1, u_2)\big|_{\textrm{(a)}}
\left(
\frac{1}{-p^+}\frac{\slashed{\bar{n}}}{2}H_\textrm{QCD}
\right)
\nonumber \\
&=&
-
ig_s^2C_F
\textrm{Tr}
\int_k
\frac{
P_n 
v(p)\bar{u}(p)
\frac{1}{p^+}
\left(
N^F
+
N_\perp^F
\right)
H_\textrm{QCD}
}
{(k^2-m^2+i\varepsilon)
[(p-k)^2+i\varepsilon]
(p^2-m^2+i\varepsilon)},
\end{eqnarray}
where
\begin{eqnarray}
N^F
&=&
m^2[Dp^+-(D-2)k^+]\frac{\slashed{{n}}\slashed{\bar{n}}}{4}
+
m[Dm^2 - (D-2)p^+k^-]\frac{\slashed{\bar{n}}}{2},
\nonumber \\
N^F_\perp
&=&
\left\{
-
Dm{p}_\perp^2
+
(D-2)\left[k^- p^+\slashed{p}_\perp - (m^2-{p}_\perp^2)
\slashed{k}_\perp
\right]
\right\}
\frac{\slashed{\bar{n}}}{2}
\nonumber \\
&&
+
\left[
Dm p^+ \slashed{p}_\perp
-
(D-2)
\left(
mp^+\slashed{k}_\perp
+p^+ \slashed{k}_\perp \slashed{p}_\perp
-
k^+ {p}_\perp^2
\right)
\right]
\frac{\slashed{n}\slashed{\bar{n}}}{4}.
\end{eqnarray}

Next, let us consider the contribution to $G(u_1,u_2)$.  Note that $G$
receives no contribution from Fig.~\ref{fig:SCETself}(a), because the
collinear gluon building block $\mathcal{G}_{n\perp\nu}^\dagger$ must
generate at least one gluon field at $s_1\bar{n}$, while the diagram in
Fig.~\ref{fig:SCETself}(a) does not contain any gluon fields at that
point.  The contribution to $G(u_1,u_2)$ from the diagram in
Fig.~\ref{fig:SCETself}(b) is given by
%---------------
\begin{eqnarray}
%---------------
\label{eq:SECTself2}%
G(u_1,u_2)\big|_{\textrm{(b)}}
%&=&
%\frac{(2p^+)^3 }{(2\pi)^3}
%\int [ds]\,
%e^{-i(s_1u_1+s_2u_2)(2p^+)}
%\int_k
%e^{ik^+s_1}
%e^{ip^+s_2}
%e^{i(p^+-k^+)s_3}
%\nonumber \\
%&&
%\times
%P_n 
%v(p)\bar{u}(p)
%(ig_s\gamma^\rho T_f^a)
%\frac{i(\slashed{k}+m)}{k^2-m^2+i\varepsilon}
%P_{\bar{n}}
%\frac{-ig_{\rho\nu}^\perp}{(p-k)^2+i\varepsilon}
%\nonumber \\
&=&
ig_s^2C_F
\delta(u_1- \tfrac{1}{2})
\delta(u_2 - \tfrac{1}{2})
\int_k
\frac{P_n 
v(p)\bar{u}(p)
\gamma_{\perp\nu}
(\slashed{k}+m)
P_{\bar{n}}\gamma_\perp^\nu
}{(k^2-m^2+i\varepsilon)[(p-k)^2+i\varepsilon]}.
%---------------
\end{eqnarray}
%---------------

The Wilson coefficient for this diagram follows from the discussion after
Eq.~(\ref{eq:exp-phin}) and is given by 
%---------------
\begin{eqnarray}
\label{eq:Wilson-for-G}
%---------------
H_{G}
&=&\frac{1}{-p^+}\frac{\slashed{\bar{n}}}{2}H_\textrm{QCD}.
%---------------
\end{eqnarray}
%---------------
Combining this with the expression for $G(u_1,u_2)\big|_{\textrm{(b)}}$ in Eq.~(\ref{eq:SECTself2}), we
    obtain
\begin{eqnarray}
\label{eq:amp_from_G}
i\mathcal{A}_{G}
&=&
{\rm Tr}\int du_1 du_2 
G(u_1, u_2)\big|_{\textrm{(b)}}
\left(
\frac{1}{-p^+}\frac{\slashed{\bar{n}}}{2}H_\textrm{QCD}
\right)
\nonumber \\
&=&
ig_s^2C_F
{\rm Tr}\int_k
\frac{P_n 
v(p)\bar{u}(p)
\frac{1}{p^+}
\left(
N^{G}+N_\perp^{G}
\right)
H_\textrm{QCD}
}{(k^2-m^2+i\varepsilon)[(p-k)^2+i\varepsilon]},
\end{eqnarray}
where
\begin{eqnarray}
N^{G}
&=&
(D-2)
k^+
\frac{\slashed{{n}}\slashed{\bar{n}}}{4}
-
(D-2)m\frac{\slashed{\bar{n}}}{2},
\nonumber \\
N^{G}_\perp
&=&
(D-4)\slashed{k}_\perp \frac{\slashed{\bar{n}}}{2}.
\end{eqnarray}

Summing Eqs.~\eqref{eq:amp_from_F} and \eqref{eq:amp_from_G} with
the contribution from Eq.~\eqref{eq:Eu1u2_a}, we obtain the
self-energy amplitudes from the modified building blocks from those
diagrams that have the explicit topology of a self energy:
%---------------
\begin{eqnarray}
\label{eq:QCD-SCET-compare_modified}
%---------------
i\mathcal{A}_E
+
i\mathcal{A}_F + i\mathcal{A}_{G}
&=&
i\mathcal{A}_\textrm{QCD}
+ig_s^2C_F
{\rm Tr}\int_k
\frac{P_n 
v(p)\bar{u}(p)
(2m-2\slashed{k}_\perp)
\frac{\slashed{\bar{n}}}{2}H_\textrm{QCD}}
{p^+(k^2-m^2+i\varepsilon)
[(p-k)^2+i\varepsilon]}.
%---------------
\end{eqnarray}
%---------------
The second term on the right side of
Eq.~\eqref{eq:QCD-SCET-compare_modified} is the mismatch between the
full-QCD self-energy amplitude and the amplitude from the diagrams that
have a self-energy topology in the case of the collinear functions that
derive from the modified building blocks.  We note that this mismatch
first appears at subleading power in the SCET expansion parameter
$\lambda$ and that there is a contribution that is proportional to the
quark mass and a contribution that survives when the quark mass is set
to zero. The contribution of the second term on the right side of
    Eq.~\eqref{eq:QCD-SCET-compare_modified} to $R$, the residue of the
    two-point function at the single-quark pole, is computed in
    Appendix~\ref{app:delta-R}.

\subsubsection{Hidden self-energy contributions in diagrams that involve
  the Wilson line}
%---------------------------------------
\begin{figure}
\centering \includegraphics[scale=0.7]{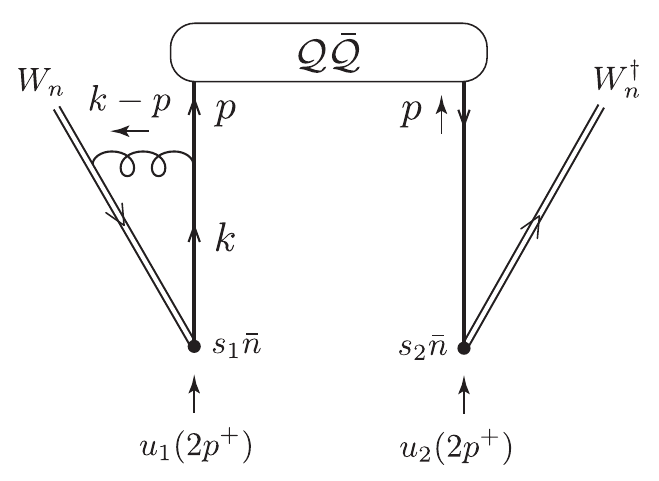}
\caption{\label{fig:SCETself4} 
The Wilson-line diagram that contributes to the quark self-energy part of $F(u_1,u_2)$.
}
\end{figure}
%--------------------------------------- 
%---------------------------------------
\begin{figure}
\centering \includegraphics[scale=0.7]{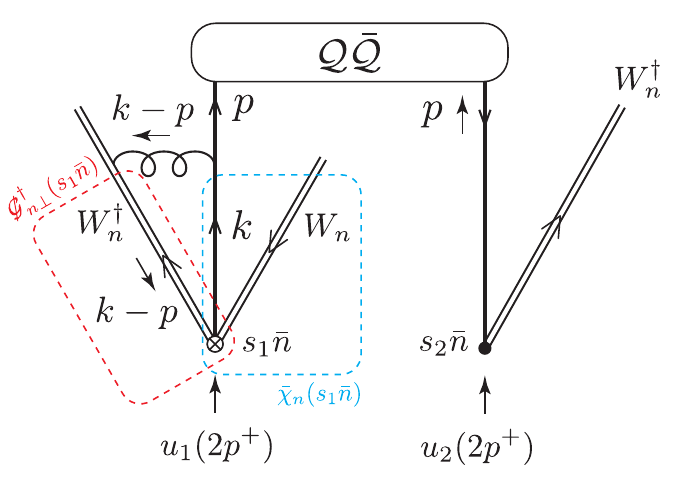}
\caption{\label{fig:SCETself5} The Wilson-line diagram that contributes
  to the quark self-energy part of $G(u_1,u_2)$. The part of the diagram
  in the cyan dashed box originates from $\bar{\chi}_n(s_1\bar{n})$,
  while the part of the diagram in the red dashed box originates from
  the longitudinally polarized gluon field that is contained in
  $\slashed{\mathcal{G}}_{n\perp}^\dagger(s_1\bar{n})$.  The crossed
  circle represents the contribution from the ordinary derivative in the
  factor $-i\overleftarrow{D}_{n\perp\nu}$ in ${\cal
    G}_{n\perp\nu}^\dagger$. This contribution yields, in momentum
  space, a factor of the negative of longitudinal-gluon momentum that
      flows into the Wilson-line, which is
  $-(\slashed{k}_\perp^\nu - \slashed{p}_\perp^\nu)$, in the present
  case.}
\end{figure}
%--------------------------------------- 

As we remarked after Eq.~\eqref{eq:exp-phin}, the use of the identities
in Eq.~(\ref{eq:Wilson-line-identity}) to obtain the modified building
blocks moves an interaction with the gluon field from the covariant
derivative in the denominator in Eq.~\eqref{eq:exp-phin} to the Wilson
line.  Hence, this re-arrangement moves self-energy contributions into
Wilson-line contributions. There are potentially two such contributions:
the contribution to $F(u_1, u_2)$ from the Wilson-line diagram in
Fig.~\ref{fig:SCETself4} and the contribution to $G(u_1,u_2)$ from
    the Wilson-line diagram in Fig.~\ref{fig:SCETself5}, which arises
    from the Wilson-line part of the building block
    $\slashed{\mathcal{G}}_{n\perp}^\dagger$ in
    Eq.~\eqref{eq:building-blocks-2}.  We now examine these hidden self-energy
contributions in detail.

First, we consider the contribution of the Wilson-line diagram in
      Fig.~\ref{fig:SCETself4} to the collinear function $F(u_1, u_2)$, which
  is defined in Eq.~\eqref{eq:F_and_G}.  This contribution is
\begin{eqnarray}
\label{eq:F-Wilson-line}%
F(u_1, u_2)\big|_{\textrm{Fig.~\ref{fig:SCETself4}}}
&=&
-ig_s^2C_F
\delta(u_1-\tfrac{1}{2}) \delta(u_2-\tfrac{1}{2})
\int_k
\frac{
P_n 
v(p)\bar{u}(p)
\slashed{\bar{n}}
(\slashed{k}+m)
P_{\bar{n}}
(\slashed{p}_\perp-m)
}
{(k^2-m^2+i\varepsilon)
[(p-k)^2+i\varepsilon]
(k^+-p^++i\varepsilon)}
\nonumber \\
&=&
-ig_s^2C_F
\delta(u_1-\tfrac{1}{2}) \delta(u_2-\tfrac{1}{2})
\int_k
\frac{
P_n 
v(p)\bar{u}(p)
\left[
2k^+(\slashed{p}_\perp - m)
\right]
P_{\bar{n}}
}
{(k^2-m^2+i\varepsilon)
[(p-k)^2+i\varepsilon]
(k^+-p^++i\varepsilon)}.
\nonumber \\
\end{eqnarray}
Here, the derivative operator in $F$, $(-i\slashed{\partial}_\perp -
m)$, reads off the external momentum $p$ for
Fig.~\ref{fig:SCETself4}, not $k$.

In addition to a
self-energy contribution, this expression contains a non-self-energy
contribution. We can identify the latter by referring to the first line
of Eq.~\eqref{eq:exp-phin}, where we see that the ordinary derivative in
the denominator acts only on the covariant derivative and the quark
field prior to the re-arrangement, while, after the re-arrangement,
the ordinary derivative in the denominator acts on the Wilson line, as
well.  Hence, the non-self-energy contribution from the diagram of
Fig.~\ref{fig:SCETself4} is obtained by multiplying the expression in
Eq.~(\ref{eq:F-Wilson-line}) by $p^+/k^+$.
The resulting expression is
\begin{eqnarray}
\label{eq:F-Wilson-line-non}%
F(u_1, u_2)\big|_{\textrm{Fig.~\ref{fig:SCETself4}}}^{\rm non-self}
&=&
-ig_s^2C_F\delta(u_1-\tfrac{1}{2}) \delta(u_2-\tfrac{1}{2})
\int_k
\frac{
P_n 
v(p)\bar{u}(p)
\left[
2p^+(\slashed{p}_\perp - m)
\right]
P_{\bar{n}}
}
{(k^2-m^2+i\varepsilon)
[(p-k)^2+i\varepsilon]
(k^+-p^++i\varepsilon)}.
\nonumber \\
\end{eqnarray}

Subtracting the expressions in Eqs.~(\ref{eq:F-Wilson-line}) and
(\ref{eq:F-Wilson-line-non}), multiplying by the Wilson coefficient in
Eq.~(\ref{eq:hard_F}), and integrating over $u_1$ and $u_2$, we find that
the amplitude for the self-energy contribution of $F(u_1, u_2)$ from
Fig.~\ref{fig:SCETself4} is
\begin{eqnarray}
\label{eq:hidden_self}
i\mathcal{A}_F\big|_{\textrm{Fig.~\ref{fig:SCETself4}}}^\textrm{self}
&\equiv&
i\mathcal{A}_F\big|_{\textrm{Fig.~\ref{fig:SCETself4}}}
-
i\mathcal{A}_{F}\big|_{\textrm{Fig.~\ref{fig:SCETself4}}}^{\rm non-self}
\nonumber \\
&=&
-ig_s^2C_F
{\rm Tr}
\int_k
\frac{
P_n 
v(p)\bar{u}(p)
\left(
2m-2\slashed{p}_\perp
\right)
\frac{\slashed{\bar{n}}}{2} H_\textrm{QCD}
}
{p^+(k^2-m^2+i\varepsilon)
[(p-k)^2+i\varepsilon]}.
\end{eqnarray}

Next, we consider the contribution of the Wilson-line diagram in
Fig.~\ref{fig:SCETself5} to the collinear function $G(u_1, u_2)$, which
is defined in Eq.~\eqref{eq:F_and_G}. This contribution arises from the
longitudinally polarized gluon field that is contained in $W_n^\dagger$
in ${\cal G}_{n\perp\nu}^\dagger$. In momentum space, ${\cal
  G}_{n\perp\nu}^\dagger$ in Eq.~\eqref{eq:building-blocks-2} can be expanded in
terms of transverse and longitudinal gluon fields as
\begin{eqnarray}
\label{eq:gluon-jet-field-expanded}
{\cal G}_{n\perp\nu}^\dagger
=
g_sG_{n\perp\nu}
-
\frac{q_{\perp\nu}}{\bar{n}\cdot q+i\varepsilon}
g_s\bar{n}\cdot G_n
+
\cdots,
\end{eqnarray}
where the ellipsis denotes terms involving more than one gluon field and
$q$ is the longitudinal-gluon momentum that flows into the Wilson line
$W_n^\dagger$.  In the second term, the factor $-q_\perp^\nu$ arises
from the ordinary derivative in the operator
$-i\overleftarrow{D}_{n\perp\nu}$ in ${\cal G}_{n\perp\nu}^\dagger$,
which reads off the negative of the longitudinal-gluon momentum that
flows into the Wilson line.  The Wilson-line propagator and vertex from
$W_n^\dagger$ in ${\cal G}_{n\perp\nu}^{\dagger}$ generate the factor
$\bar{n}\cdot G_n/(\bar{n}\cdot q+i\varepsilon)$.\footnote{Note that the
active Wilson line in Fig.~\ref{fig:SCETself5} is associated with
$W_n^\dagger$ inside $\slashed{\mathcal{G}}_{n\perp}^\dagger$, as
defined in Eq.~\eqref{eq:building-blocks-2}, rather than with $W_n$ in
$\bar{\chi}_n$, as defined in Eq.~\eqref{eq:building-blocks-1}.}  Hence,
the contribution of Fig.~\ref{fig:SCETself5} to the amplitude that
arises from $G(u_1,u_2)$ is
\begin{eqnarray}
\label{eq:amp_from_Gwilsonline}
i\mathcal{A}_{G}\big|_{\textrm{Fig.~\ref{fig:SCETself5}}}=
ig_s^2C_F
{\rm Tr}\int_k
\frac{P_n 
v(p)\bar{u}(p)
\frac{k^+}{p^+}
\left(
2\slashed{k}_{\perp}-2\slashed{p}_{\perp}
\right)
\frac{\slashed{\bar{n}}}{2} H_\textrm{QCD}
}{(k^2-m^2+i\varepsilon)[(p-k)^2+i\varepsilon]\left(k^+-p^++i\varepsilon\right)},
\end{eqnarray}
where we have combined the contribution to $G(u_1, u_2)$ from
this diagram with the Wilson coefficient for $G(u_1, u_2)$ in
Eq.~\eqref{eq:Wilson-for-G}. 

As in the analysis of 
$i\mathcal{A}_F\big|_{\textrm{Fig.~\ref{fig:SCETself4}}}$, we extract
the non-self-energy part of $i\mathcal{A}_{G}\big|_{\textrm{Fig.~\ref{fig:SCETself5}}}$ by multiplying the integrand by
$p^+/k^+$. This yields
\begin{eqnarray}
\label{eq:amp_from_Gwilsonlinenonself}
i\mathcal{A}_{G}\big|_{\textrm{Fig.~\ref{fig:SCETself5}}}^{\text{non-self}}
=
ig_s^2C_F
{\rm Tr}\int_k
\frac{P_n 
v(p)\bar{u}(p)
\left(
2\slashed{k}_{\perp}-2\slashed{p}_{\perp}
\right)
\frac{\slashed{\bar{n}}}{2} H_\textrm{QCD}
}{(k^2-m^2+i\varepsilon)[(p-k)^2+i\varepsilon]\left(k^+-p^++i\varepsilon\right)}.
\end{eqnarray}
Subtracting the non-self-energy contribution from the full expression,
we obtain the self-energy contribution
\begin{eqnarray}
\label{eq:amp_from_Gwilsonlineself}
i\mathcal{A}_{G}\big|_{\textrm{Fig.~\ref{fig:SCETself5}}}^{\text{self}}
=
ig_s^2C_F
{\rm Tr}\int_k
\frac{P_n 
v(p)\bar{u}(p)
\left(
2\slashed{k}_{\perp}-2\slashed{p}_{\perp}
\right)
\frac{\slashed{\bar{n}}}{2} H_\textrm{QCD}
}{p^+(k^2-m^2+i\varepsilon)[(p-k)^2+i\varepsilon]}.
\end{eqnarray}

Combining this result with the self-energy contribution in
Eq.~\eqref{eq:hidden_self}, we obtain
\begin{eqnarray}
\label{eq:hiddenselftotal}
i\mathcal{A}_F\big|_{\textrm{Fig.~\ref{fig:SCETself4}}}^\textrm{self}+
i\mathcal{A}_{G}\big|_{\textrm{Fig.~\ref{fig:SCETself5}}}^{\text{self}}
=
-ig_s^2C_F
{\rm Tr}
\int_k
\frac{
P_n 
v(p)\bar{u}(p)
\left(
2m-2\slashed{k}_\perp
\right)
\frac{\slashed{\bar{n}}}{2} H_\textrm{QCD}
}
{p^+(k^2-m^2+i\varepsilon)
[(p-k)^2+i\varepsilon]}.
\end{eqnarray}
This contribution is proportional to the factor
$(\slashed{k}_{\perp}-m)\frac{\slashed{\bar{n}}}{2}$, which is expected,
since the operator in the second term of the last equality in the
transformation in Eq.~(\ref{eq:exp-phinb}) is proportional to
$\left(-i\overleftarrow{\slashed{\partial}}_\perp -m\right)
\frac{\slashed{\bar n}}{2}$.  The contribution in
Eq.~(\ref{eq:hiddenselftotal}) accounts for the mismatch in
Eq.~\eqref{eq:QCD-SCET-compare_modified}, and we find that
%---------------
\begin{eqnarray}
%---------------
i\mathcal{A}_E
+
i\mathcal{A}_F + i\mathcal{A}_G
+
i\mathcal{A}_F\big|_{\textrm{Fig.~\ref{fig:SCETself4}}}^\textrm{self}+
i\mathcal{A}_{G}\big|_{\textrm{Fig.~\ref{fig:SCETself5}}}^{\text{self}}
&=&
i\mathcal{A}_\textrm{QCD}.
%---------------
\end{eqnarray}
%---------------

From this analysis, we see that, for the operator basis that is
constructed from the building blocks in Eq.~(\ref{eq:BB-modified}),
self-energy contributions can be hidden in diagrams involving Wilson
lines. Furthermore, the relevant diagram also contains a non-self-energy
contribution, and so, the process of identifying the self-energy
contribution is somewhat subtle.  The implications of this subtlety for
the construction of the $S$-matrix are discussed in
Sec.~\ref{sec:complete-LSZ}.

\section{Complete LSZ Formula \label{sec:complete-LSZ}}

As we have mentioned, the complete LSZ formula contains, in addition to
the amputated Green's function, a factor $\sqrt{R}$ for each external
leg, where $R$ is the residue of the two-point function at the
single-quark pole.\footnote{Examples of the application of the LSZ
formula in full QED and in SCET can be found in
Refs.~\cite{vanBijleveld:2025ekz,Fleming:2007xt}.}

In the standard LSZ formula, if $\phi$ is the quark field that
appears in the Green's function (not to be confused with the SCET field
$\phi_n$), then, for an external leg with momentum $p$,
\begin{eqnarray}
R&=&\lim_{\slashed{p}\to m} -i(\slashed{p}-m)\int d^4x\, e^{ip\cdot x} 
\langle 0\vert T\phi(x)\bar\phi(0)\vert 0\rangle.
\end{eqnarray}
The factor to be amputated on an outgoing external leg with momentum $p$ is 
\begin{subequations}
\label{eq:amp-factor}
\begin{eqnarray}
&&\frac{iZ_{\rm out}}{\slashed{p}-m},
\end{eqnarray}
and the factor to be amputated from an incoming external leg with momentum $p$ is
\begin{eqnarray}
&&\frac{iZ_{\rm in}}{\slashed{p}-m}.
\end{eqnarray}
\end{subequations}
In the standard LSZ formula,
\begin{eqnarray}
Z_{\rm out}&=&Z_{\rm in}=R.
\end{eqnarray}

Neither the amputated Green's function, nor the factors $\sqrt{R}$ are
gauge invariant.  However, as we have stressed, the complete LSZ formula
is gauge invariant, as is shown, for example, on pp.~344--346 of
Ref.~\cite{Sterman:1993hfp}.  As we will see, it is convenient to
consider a factor $R$ for each pair of incoming-outgoing external
fermion legs, rather than a factor $\sqrt{R}$ for each external leg.

\subsection{Generalization of the LSZ formula}

There is some freedom to move contributions from $Z_{\rm out}$ and
$Z_{\rm in}$ to $R$, since the $S$-matrix is unchanged provided that the
ratio $R/(Z_{\rm out} Z_{\rm in})$ is unchanged.  
We wish to exploit this freedom in the discussions
below.

We can implement this freedom formally by introducing an ``amputating
field'' $\zeta$ that determines the factor to be amputated from the
Green's function.  Like $\phi$, $\zeta$ is an interpolating field.  That
is, its matrix element between the vacuum and an in or out state is
nonzero. Now, we write
\begin{subequations}
\label{eq:LSZ-factors}%
\begin{eqnarray}
Z_{\rm out}&=&\lim_{\slashed{p}\to m} -i(\slashed{p}-m)\int d^4x\, e^{ip\cdot x} 
\langle 0\vert T\phi(x)\bar\zeta(0)\vert 0\rangle,\\
Z_{\rm in}&=&\lim_{\slashed{p}\to m} -i(\slashed{p}-m)\int d^4x\, e^{ip\cdot x} 
\langle 0\vert T\zeta(x)\bar\phi(0)\vert 0\rangle.
\end{eqnarray}
Then, we must take
\begin{eqnarray}
R&=&\lim_{\slashed{p}\to m} -i(\slashed{p}-m)\int d^4x\, e^{ip\cdot x} 
\langle 0\vert T\zeta(x)\bar\zeta(0)\vert 0\rangle.
\end{eqnarray}
\end{subequations}
It follows from the discussion on p.~99 of Ref.~\cite{Sterman:1993hfp}, that
\begin{eqnarray}
Z_{\rm out}&=&(2\pi)^3 \langle 0\vert \phi(0)\vert \bm{p}\rangle
\langle \bm{p}\vert \bar\zeta(0)\vert 0 \rangle,
\nonumber\\
Z_{\rm in}&=&(2\pi)^3 \langle 0\vert \zeta(0)\vert \bm{p}\rangle
\langle \bm{p}\vert \bar\phi(0)\vert 0\rangle,
\nonumber\\
R&=&(2\pi)^3 \langle 0\vert \zeta(0)\vert \bm{p}\rangle
\langle \bm{p}\vert \bar\zeta(0)\vert 0\rangle,
\end{eqnarray}
where $\vert \bm{p}\rangle$ is a single-particle state of 3-momentum
$\bm{p}$.
Hence, the ratio 
\begin{eqnarray}
\frac{R}{Z_{\rm out} Z_{\rm in}}&=&\frac{1}{(2\pi)^{3}
\langle 0\vert \phi(0)\vert \bm{p}\rangle
\langle \bm{p}\vert \bar\phi(0)\vert 0\rangle}, 
\end{eqnarray}
and, consequently, the $S$-matrix, are independent of the choice of $\zeta$. 

\subsection{LSZ formula in full QCD}

\begin{figure}
\centering \includegraphics[scale=0.7]{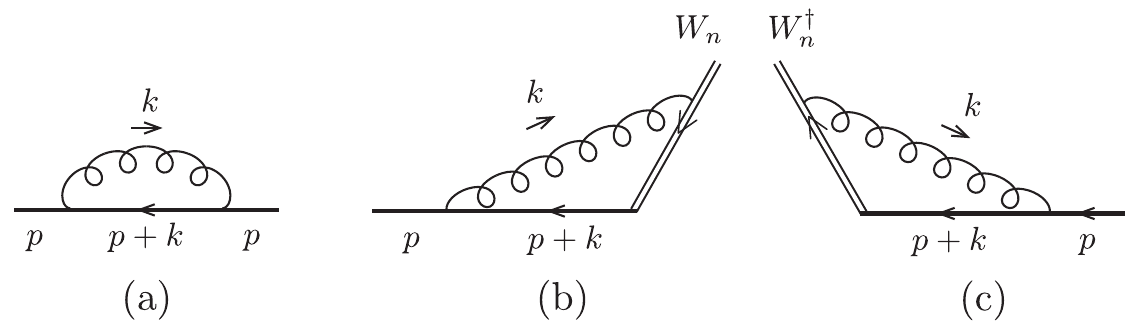}
\caption{\label{fig:two-point} Contributions to the
  two-point function at the one-loop level.}
\end{figure}

Let us first recall the construction of the LSZ formula
in full QCD. The standard choice for the amputating
field $\zeta$ is $\psi$. 
In this case,
\begin{eqnarray}
\label{eq:full-QCD-LSZ-1}%
Z_{\rm out}^{\rm QCD}&=&Z_{\rm in}^{\rm QCD}=R^{\rm QCD}=\lim_{\slashed{p}\to m} -i(\slashed{p}-m)\int d^4x\, e^{ip\cdot x} 
\langle 0\vert T\psi(x)\bar\psi(0)\vert 0\rangle.
\end{eqnarray}
At the one-loop level, the two-point
function contains diagram (a) in Fig.~\ref{fig:two-point}, and only
contributions involving diagram (a) are amputated from the Green's
function.

Our calculations so far have been in the Feynman gauge. Of course, the
$S$-matrix is gauge invariant, and so we can compute it in any gauge. In
particular, we can compute the amputated Green's function and the
two-point function in $R$ in the light-cone gauge.  As we discuss, in
Appendix~\ref{app:light-cone-gauge}, the full-QCD, direct-QCD-basis, and
modified-basis self-energy contributions are all equal in the light-cone
gauge. We call the terms in the polarization sum in the
    light-cone-gauge gluon propagator that are proportional to the gluon
    momentum the ``gauge terms.'' We show that the contributions of the
    gauge terms are equivalent
    to the Wilson-line contributions in the Feynman gauge. We note,
however, that it is preferable to calculate in a covariant gauge, as the
light-cone-gauge expressions have ambiguities in the treatment of the
denominator of the gluon-propagator gauge term that are only resolved by
referring to a covariant-gauge calculation.
 
\subsection{LSZ formula in the direct-QCD basis}

Now let us consider the construction of the LSZ formula in the
direct-QCD basis.

One option would be to amputate only contributions on external legs
    that have the topological form of a self-energy contribution. This
    corresponds to taking the amputating field $\zeta$ in
Eq.~\eqref{eq:LSZ-factors} to be a linear combination of the charge
conjugates of fields that appear at the point $s_1\bar n$ in the
definitions of the collinear functions $E$ and $J$ in
Eqs.~(\ref{eq:collinear-fn-E}) and (\ref{eq:collinear-fn-J}), weighted
according to their respective short-distance coefficients, but with
$W_n^\dagger$ set equal to unity. That is, we take the amputating field
to be
\begin{eqnarray}
\zeta^{{\rm dir-no-}W}&=&\xi_n-\frac{\slashed{\bar{n}}}{2}
\frac{1}{i\bar{n}\cdot D_n}
(i\slashed{D}_{n\perp} - m) \xi_n.
\end{eqnarray}
As we have seen from the analysis in Sec.~\ref{subsec:direct-QCD},
with this option, $Z_{\rm out}$ and $Z_{\rm in}$ are equal and are given
by the full-QCD expression in Eq.~(\ref{eq:full-QCD-LSZ-1}). It follows
from the invariance of the $S$-matrix under changes of amputating field
that $R$ is also given by the full-QCD expression in
Eq.~(\ref{eq:full-QCD-LSZ-1}).  Indeed, using Eqs.~\eqref{eq:EOM} and
    (\ref{eq:projectors}), we can write
\begin{eqnarray}
\zeta^{{\rm dir-no-}W}&=&\xi_n+\eta_n=\psi.
\end{eqnarray}
Hence, this choice corresponds to
amputating only diagram (a) of Fig.~\ref{fig:two-point}, and the
two-point function in $R$ contains only diagram (a).

Another option would be to amputate all of the Wilson-line
    contributions on external legs, along with the contributions on
    external legs that have self-energy topologies.  This corresponds
    to taking the amputating field $\zeta$ in
Eq.~\eqref{eq:LSZ-factors} to be a linear combination of the charge
conjugates of fields that appear at the point $s_1\bar n$ in the
definitions of the collinear functions $E$ and $J$ in
Eqs.~(\ref{eq:collinear-fn-E}) and (\ref{eq:collinear-fn-J}), weighted
according to their respective short-distance coefficients, but now
retaining the factors $W_n^\dagger$.  That is, we take the amputating
field to be
\begin{eqnarray}
\zeta^{{\rm dir-}W}&=&W_n^\dagger\xi_n-W_n^\dagger
\frac{\slashed{\bar{n}}}{2}
\frac{1}{i\bar{n}\cdot D_n}
(i\slashed{D}_{n\perp} - m)\xi_n.
\end{eqnarray}
Using Eqs.~(\ref{eq:building-blocks-1}) and (\ref{eq:projectors}), we can write
\begin{eqnarray}
\zeta^{{\rm dir-}W}&=&W_n^\dagger \xi_n+W_n^\dagger\eta_n=W_n^\dagger\psi.
\end{eqnarray}
Hence, the matrix element in $Z_{\rm out}$ contains both diagrams (a)
and (b) in Fig.~\ref{fig:two-point}, and contributions involving these
diagrams on an outgoing external leg are to be amputated from the
Green's function. Similarly, the matrix element in $Z_{\rm in}$ contains
both diagrams (a) and (c) in Fig.~\ref{fig:two-point}, and contributions
involving these diagrams on an incoming external leg are to be amputated
from the Green's function.  The two-point function in $R$ contains all
three diagrams in Fig.~\ref{fig:two-point}.

We see that the Wilson-line contributions that are amputated from the
Green's function are now contained in the two-point function in $R$. We
conclude that we can move Wilson-line contributions freely between the
truncated Green's function and $R$ without changing the $S$-matrix.
This is not surprising, since Wilson lines arise from field
redefinitions that absorb pure gauge interactions, and the $S$-matrix is
gauge invariant.  As we will see in Appendix~\ref{app:light-cone-gauge},
this construction of the LSZ formula, in which the Wilson-line
contributions are in $R$, corresponds to the construction in the
light-cone gauge. However, as we have already mentioned, it is
preferable to calculate in a covariant gauge, as the light-cone-gauge
expressions have ambiguities in the treatment of the denominator of the
gluon-propagator gauge term that are only resolved by referring to a
covariant-gauge calculation.

\subsection{LSZ formula in the modified basis}

Let us now consider options for the construction of the LSZ formula in
the case of the modified building blocks.

One option would be to amputate as part of external legs only the
diagrams that have a self-energy topology and leave all external-leg
Wilson-line contributions in the truncated Green's function.  This
option corresponds to taking the amputating field $\zeta$ to be a linear
combination of the charge conjugates of the fields that appear at
the point $s_1\bar n$ in the definitions of the collinear functions $E$,
$F$, and $G$ in Eqs.~(\ref{eq:collinear-fn-E}), (\ref{eq:F_and_G}),
weighted by their respective short-distance coefficients, but with
$W_n^\dagger$ set to unity. That is, we take the amputating field to be
\begin{eqnarray}
\zeta^{{\rm mod-no-}W}
&=&\xi_n
-\frac{\slashed{\bar{n}}}{2} \frac{1}{i\bar n\cdot \partial}
g_s \slashed{G}_{n\perp} \xi_n
-\frac{\slashed{\bar{n}}}{2}\frac{1}{i\bar n \cdot\partial}
\left(i\slashed{\partial}_\perp-m\right) \xi_n.
\end{eqnarray}
However, as we have seen in
Sec.~\ref{sec:calc-mod-blocks}, this procedure does not lead to
expressions for $Z_{\rm out}$, $Z_{\rm in}$, or $R$ that correspond to
the ones in full QCD. The reason for this is that, in the modified
basis, part of the external self-energy contribution is hidden in
Wilson-line diagrams and is, therefore, not selected by a prescription
that is based only on self-energy topology.  We can see this directly
    by using Eqs.~(\ref{eq:building-blocks-1}) and (\ref{eq:projectors})
    to write
\begin{eqnarray}
\zeta^{{\rm mod-no-}W}
&=&\xi_n+\eta_n-\frac{\slashed{\bar{n}}}{2}\bigg(\frac{1}{i\bar{n}\cdot\partial}-
\frac{1}{i\bar{n}\cdot D_n}\bigg)(i\slashed{D}_{ n\perp}-m)\xi_n\nonumber\\
&=&\psi-\frac{\slashed{\bar{n}}}{2}\bigg(\frac{1}{i\bar{n}\cdot\partial}-
W_n \frac{1}{i\bar{n}\cdot \partial}W_n^\dagger\bigg)
(i\slashed{D}_{n\perp }-m)
\xi_n.
\end{eqnarray}
In each expression after the equals signs, the field $\xi_n+\eta_n=\psi$
generates the full-QCD expressions for $Z_{\rm out}$, $Z_{\rm in}$, and
$R$, and the remaining term generates the hidden self-energy
contributions.

Although a particular choice of amputating field need not yield
individual factors that are equal to their full-QCD counterparts, the
present topology-based separation is inconvenient because the isolated
hidden self-energy contribution to $R$ has an ill-defined mass-shell
limit (Appendix~\ref{app:delta-R}). As we have seen, in the modified
    basis, there is no simple
diagrammatic prescription for assigning the hidden self-energy
contribution separately to the propagators on external legs.

Another option in the case of the modified basis would be to
amputate as part of the external legs all external-leg Wilson-line
contributions, as well as contributions that have a self-energy
topology. This option corresponds to taking the amputating field $\zeta$
to be a linear combination of the charge-conjugates of the fields that
appear at the point $s_1\bar n$ in the definitions of the collinear
functions $E$, $F$, and $G$ in Eqs.~(\ref{eq:collinear-fn-E}) and
(\ref{eq:F_and_G}), weighted by their respective short-distance
coefficients. That is, we take the amputating field to be
\begin{eqnarray}
\zeta^{{\rm mod-}W}
&=&\chi_n
-\frac{\slashed{\bar{n}}}{2} \frac{1}{i\bar n\cdot \partial}
\slashed{\cal G}_{n\perp} W^\dagger_n\xi_n
-\frac{\slashed{\bar{n}}}{2}\frac{1}{i\bar n \cdot\partial}
\left(i\slashed{\partial}_\perp-m\right) W^\dagger_n\xi_n.
\end{eqnarray}
Using Eqs.~(\ref{eq:exp-phin}) and (\ref{eq:projectors}), we can write
\begin{eqnarray}
\label{eq:zeta-mod-W}
\zeta^{{\rm mod-}W}
&=&\chi_n+\phi_n=W_n^\dagger \psi.
\end{eqnarray}
Hence, we have the same amputating field as in the corresponding option
for the direct-QCD basis. Obviously, the resulting expressions for
$Z_{\rm out}$, $Z_{\rm in}$, and $R$ are equal to those for the
direct-QCD basis, and, as is shown in
Appendix~\ref{app:light-cone-gauge}, they are equivalent to the
expressions in the light-cone gauge.  Again, all of the Wilson-line
contributions (including the hidden self-energy contributions) reside in
$R$, and none reside in the truncated Green's function.

\subsection{Renormalization}

We note that, in full QCD in the on-shell renormalization scheme for
    massive quarks, the quark wave-function-renormalization constant
    $Z_{\psi}$ is usually chosen so as to absorb all of the
    contributions to $R$. For the options in SCET for the construction
    of the LSZ formula that lead to the full-QCD expression for $R$,
    there is no particular difficulty in implementing this
    convention. However, in the case of the modified basis in the
    Feynman gauge, the option in which only diagrams with a self-energy
    topology are amputated from the Green's function leads to an
    expression for $R$ that differs from that in full QCD. Therefore, there
    would, after renormalization, be a remainder of $R$ that must be
    taken into account in order to obtain a correct $S$-matrix.  (In
    the case of massless quarks, the remainder in $R$ for this
    modified-basis option would differ from the remainder in $R$ in full
    QCD.) This remainder corresponds to the hidden self-energy
    contribution. As we have mentioned, and is shown in
    Appendix~\ref{app:delta-R}, the hidden self-energy contribution to
    $R$ is not, by itself, well defined.  As we have also emphasized, in
    the case of the modified basis in the Feynman gauge, there is no
    simple diagrammatic prescription for the construction of the LSZ
    formula that leads to an expression for $R$ that is equal to the
    full-QCD expression in the Feynman gauge.

\section{Summary\label{sec:summary}}

In this paper, we have examined quark self-energy contributions in
SCET. This examination was motivated by the requirement, in constructing
the $S$-matrix through the LSZ reduction formula, to identify and
amputate complete propagators on the external legs of a Green's function
and to evaluate the complete two-point function on the mass shell.  The
complete propagators are characterized in full QCD by 1PI self-energy
contributions that alternate with free-particle propagators. In full
QCD, it is simple to identify self-energy contributions and, hence, the
complete propagators, from their topologies.  We have found that, for a
SCET operator basis that follows directly from QCD (the direct-QCD
    basis), diagrams that contribute to quark self energies can be
identified from their topologies, as in full QCD.  However, for an
alternative operator basis that follows from the use of a Wilson-line
    identity (the modified basis), self-energy contributions appear in
diagrams that involve Wilson lines. This phenomenon first occurs at
subleading power in the SCET expansion parameter $\lambda$. There is a
contribution that is proportional to the quark mass and a contribution
that is nonzero even in the massless case. The identification of these
hidden self-energy contributions is subtle because the relevant diagrams
that involve Wilson lines also contain non-self-energy contributions.
Furthermore, the hidden self-energy contributions to the two-point
    function are ill-defined in the mass-shell limit, making their
    computation as isolated quantities problematic.

We have introduced a generalization of the LSZ reduction formula in
which one can make different choices for the complete propagator that is
to be amputated from the Green's function.  We show that this
generalization of the LSZ formula compensates for different choices for
the complete propagator through the factor that arises from the on-shell
residue of the two-point function. We have used this generalization
of the LSZ formula to explore various options for the construction of the
$S$-matrix in the two SCET operator bases that we have investigated in
this paper.

Our principal analysis was carried out in the Feynman
gauge. However, we expect the problem of hidden self-energy
contributions to be present generally in other covariant gauges. In the
light-cone gauge $\bar{n}\cdot G=0$, we have found that the problem of
hidden self-energy contributions does not arise because the Wilson lines
are equal to unity and the $\bar{n}\cdot G$ contributions in the
covariant derivatives in the first lines of Eqs.~(\ref{eq:exp-phin}a)
and (\ref{eq:exp-phin}b) vanish. We have shown that the light-cone gauge
calculation is equivalent to one in which all of the Wilson-line
contributions are amputated from the Green's function as part of the
complete propagator and are included, instead, in the two-point
function. However, the light-cone-gauge calculation contains ambiguities
that arise from the denominator of the polarization sum in the gluon
propagator. Those ambiguities are resolved only through reference to a
covariant-gauge calculation.

The difficulty that we have identified originates in a specific
operation---the use of Wilson-line identities to trade an interaction in
a covariant derivative for an interaction with a Wilson line---and we,
therefore, expect it to arise in any SCET basis that is obtained through
such a re-arrangement. The modified basis can also be obtained in a
    bottom-up construction that relies on symmetry arguments alone. The
    bottom-up construction obscures the field redefinitions that are
    responsible for the hidden self-energy contributions. Although we
    have not investigated the hidden self-energy mechanism in effective
    field theories other than SCET, the nature of the mechanism suggests
    that analogous difficulties could arise in the construction of the
    $S$-matrix in other gauge effective field theories in which similar
    field redefinitions occur, either explicitly (in top-down
    construction) or implicitly (in bottom-up construction).

% If you have acknowledgments, this puts in the proper section head.
%%%%%%%%%%%%%%%%%%%%%%%%%%%%%%%%%%%%%%%%%%%%%%%%%%%%%%%%%%%%%%
\begin{acknowledgments}
%%%%%%%%%%%%%%%%%%%%%%%%%%%%%%%%%%%%%%%%%%%%%%%%%%%%%%%%%%%%%%
We wish to thank Jungil Lee, Hee Sok Chung, Patrick Hager, Jian
Wang, and Yunlu Wang for helpful discussions.  The work of G.T.B.\ is
supported by the U.S.\ Department of Energy, Division of High Energy
Physics, under Contract No.\ DE-AC02-06CH11357.
The work of D.K.\ is supported by the National Natural Science Foundation of China (NSFC) through National Key Research and Development Program under the contract No.~2024YFA1610503.
The work of X.-P.~W.\  is supported by the National Natural
Science Foundation of China under Grant
No.~12135006. The submitted
manuscript has been created in part by UChicago Argonne, LLC, Operator
of Argonne National Laboratory. Argonne, a U.S.\ Department of Energy
Office of Science laboratory, is operated under Contract
No.\ DE-AC02y-06CH11357. The U.S.\ Government retains for itself, and
others acting on its behalf, a paid-up nonexclusive, irrevocable
worldwide license in said article to reproduce, prepare derivative
works, distribute copies to the public, and perform publicly and display
publicly, by or on behalf of the Government.

%%%%%%%%%%%%%%%%%%%%%%%%%%%%%%%%%%%%%%%%%%%%%%%%%%%%%%%%%%%%%%
\end{acknowledgments}
%%%%%%%%%%%%%%%%%%%%%%%%%%%%%%%%%%%%%%%%%%%%%%%%%%%%%%%%%%%%%%

% Specify following sections are appendices. Use \appendix* if there
% only one appendix.
% \appendix*
\appendix

\section{Computation of $\delta R$\label{app:delta-R}}

In this appendix, we compute $\delta R$, the contribution to $R$ that
arises from the second term on the right side of
Eq.~\eqref{eq:QCD-SCET-compare_modified}.

$R$ is given by
\begin{eqnarray}
\label{eq:R}
R^{-1}&=&1-[\hbox{the coefficient of $(\slashed{p}-m)$ in an expansion
    of $\Sigma(p)$ about $\slashed{p}=m$}],
\end{eqnarray}
where $\Sigma(p)$ is the 1PI self-energy contribution.  The contribution
to the quark self-energy from the second term in
Eq.~\eqref{eq:QCD-SCET-compare_modified} is
\begin{eqnarray}
\delta\Sigma(p)&=& -i(\slashed{p}-m)(-g_s^2)C_F \bigg[
\frac{\mu^2\exp(\gamma_{\rm E})}{4\pi}\bigg]^\epsilon \int\frac{d^D k}{(2\pi)^D}
\frac{(2m-2\slashed{k}_\perp)\frac{\slashed{\bar n}}{2}}
     {p^+(k^2-m^2+i\varepsilon) [(p-k)^2+i\varepsilon]},\nonumber\\
\end{eqnarray}
where the factor $-i(\slashed{p}-m)$ cancels the explicit factor of the
free-quark propagator that we have included in our calculations, and we
have inserted the standard $\overline{\rm MS}$ factor $ [\exp(\gamma_{\rm
  E})/(4\pi)]^\epsilon$.

A straightforward calculation of the integral yields the result
\begin{eqnarray}
\delta\Sigma(p)&=& (\slashed{p}-m)\frac{-\alpha_s C_F}{4\pi
  p^+}\bigg\{\frac{1}{\epsilon}(2m-\slashed{p}_\perp)
+2m\bigg[2+\frac{m^2}{p^2}\log\frac{m^2-p^2-i\varepsilon}{m^2}
  -\log\frac{m^2-p^2-i\varepsilon}{\mu^2}\bigg]
\nonumber\\ &&-\slashed{p}_\perp\bigg[\frac{m^2}{p^2}+2
  +\frac{m^4}{p^4}\log\frac{m^2-p^2-i\varepsilon}{m^2}
  -\log\frac{m^2-p^2-i\varepsilon}{\mu^2}\bigg]\bigg\}\frac{\slashed{\bar
    n}}{2}+\cal{O}(\epsilon).
\end{eqnarray}

Making use of Eq.~\eqref{eq:R}, we obtain, through order $\alpha_s$,
\begin{eqnarray}
\label{eq:delta-R}%
\delta R&=&\lim_{\slashed{p}\to m}
\frac{-\alpha_s C_F}{4\pi}
\frac{1}{p^+}\bigg[\frac{2m-\slashed{p}_\perp}{\epsilon}
+4m-3\slashed{p}_\perp
+(2m-\slashed{p}_\perp)\log\frac{\mu^2}{m^2}\bigg]\frac{\slashed{\bar n}}{2}+\cal{O}(\epsilon).
\end{eqnarray}
We note that the limit as $\slashed{p}\to m$ in the expression for
      $\delta{R}$ is ill-defined because $\delta\Sigma(p)$ 
      does not depend on $p$ only through $\slashed{p}$. This is in
      contrast with the situation for $\Sigma(p)$ in, for example, full
      QCD.

\section{Calculations in the light-cone gauge}
\label{app:light-cone-gauge}

In this appendix, we repeat the self-energy calculations of the main
text in the light-cone gauge. We establish that the light-cone-gauge
    contributions to the self-energies and, hence, to $R$ are identical
    in the direct-QCD building-block basis, in the modified
    building-block basis, and in full QCD. We also show the
      correspondence between the contributions to the quark self energy
      that arise from the gauge terms in the light-cone-gauge gluon
      propagator and contributions from the Wilson-line diagrams in the
      Feynman gauge.
 
The light-cone gauge condition is
\begin{equation}
\label{eq:lc-cond}
\bar{n}\cdot G = G^+ = 0.
\end{equation}
It follows that the gluon propagator in the light-cone gauge is
\begin{equation}
\label{eq:D-lc}
D^{\mu\nu}_\text{LC}(q)
= \frac{i}{q^2+i\varepsilon}
\left(
  -g^{\mu\nu}
  + \frac{\bar{n}^\mu q^\nu + q^\mu \bar{n}^\nu}{\bar{n}\cdot q}
\right).
\end{equation}
We decompose the LC propagator as
\begin{equation}
\label{eq:decomp}
D^{\mu\nu}_\text{LC}(q)
= D^{\mu\nu}_\text{FG}(q) + \delta D^{\mu\nu}(q),
\end{equation}
where $D^{\mu\nu}_\text{FG}(q) = -ig^{\mu\nu}/(q^2+i\varepsilon)$ and
\begin{equation}
\label{eq:deltaD}
\delta D^{\mu\nu}(q)
= \frac{i}{q^2+i\varepsilon}
  \left(\frac{\bar{n}^\mu q^\nu + q^\mu \bar{n}^\nu}{\bar{n}\cdot q}\right).
\end{equation}
Here, and throughout this appendix, the superscript ``FG'' denotes a
Feynman-gauge quantity, ``LC'' denotes a light-cone gauge quantity, and
$\delta$ denotes the difference between the light-cone-gauge quantity
and the Feynman-gauge quantity [$\delta(\cdot) \equiv (\cdot)^\text{LC}
  - (\cdot)^\text{FG}$].  As we have mentioned, the contributions to the
quark self energy that arise from $\delta D^{\mu\nu}$ correspond to
contributions from Wilson-line diagrams in the Feynman gauge. We will
demonstrate this point in Appendix~\ref{sec:correspondence}.

Under the gauge condition~\eqref{eq:lc-cond}, the collinear Wilson line
$W_n = P\exp[ig_s\int_{-\infty}^0 ds\,\bar{n}\cdot G_n(x+s\bar{n})] = 1$.
The building blocks in Eq.~\eqref{eq:building-blocks-1} and
Eq.~\eqref{eq:exp-phin} then both reduce to
\begin{subequations}
\label{eq:bb-lc}
\begin{align}
\chi_n &= \xi_n, \quad \bar\chi_n = \bar\xi_n, \\[4pt]
\phi_n &= \eta_n
  = -\frac{\slnbar}{2}\,\frac{1}{i\bar{n}\cdot\partial}
    (i\slashed{D}_{n\perp}-m)\xi_n,
    \label{eq:phi-lc}\\[4pt]
\bar\phi_n &= \bar\eta_n
  = \bar\xi_n(-i\overleftarrow{\slashed{D}}_{n\perp}-m)
    \frac{1}{i\bar{n}\cdot\overleftarrow{\partial}}\frac{\slnbar}{2},
    \label{eq:phibar-lc}\\[4pt]
\mathcal{G}_{n\perp\nu} &= g_s G_{n\perp\nu},
  \label{eq:cG-lc}
\end{align}
\end{subequations}
where, crucially, $i\bar{n}\cdot D_n =
i\bar{n}\cdot\partial$ (since $\bar{n}\cdot G_n = 0$), while the
transverse gluon field $G_{n\perp}$ remains nonzero.  From this, we
      see that, in the light-cone gauge, the two building-block bases
      are identical. Furthermore, since the Wilson lines are all unity,
      there are no Wilson-line diagrams of the types in
      Figs.~\ref{fig:SCETself4} and~\ref{fig:SCETself5}.  Hence,
            there are no hidden self-energy contributions in the
            light-cone gauge.

\subsection{Full QCD in the light-cone gauge and the Feynman gauge}
The full-QCD self-energy amplitude corresponding to Fig.~\ref{fig:QCDself} is
\begin{equation}
\label{eq:AmpQCD}
i\mathcal{A}_\text{QCD}^\text{gauge}
= g_s^2 C_F\,\text{Tr}\int_k
  \frac{v(p)\,\bar{u}(p)\,
  \gamma^\mu(\slashed{k}+m)\gamma^\nu
  \, D^\text{gauge}_{\mu\nu}(p-k)\,
  (\slashed{p}+m)\,H_\text{QCD}}
  {(k^2-m^2+i\eps)(p^2-m^2+i\eps)},
\end{equation}
where the superscript ``gauge'' is either FG or LC.  Using
Eq.~\eqref{eq:decomp}, we see that
\begin{equation}
\label{eq:AmpQCD-split}
i\mathcal{A}_\text{QCD}^\text{LC}
= i\mathcal{A}_\text{QCD}^\text{FG}
+ i\mathcal{A}_\text{QCD}^{\delta D},
\end{equation}
where $i\mathcal{A}_\text{QCD}^\text{FG}$ is given in Eq.~\eqref{eq:QCD-self} and $i\mathcal{A}_\text{QCD}^{\delta D}$ is given by
\begin{equation}
\label{eq:delta_QCD}
i\mathcal{A}_\text{QCD}^{\delta D}
= ig_s^2 C_F\,\text{Tr}\int_k
  \frac{v(p)\,\bar{u}(p)\,
  N_{\delta D}\,
  (\slashed{p}+m)
  H_\text{QCD}}
  {(k^2-m^2+i\eps)
  [(p-k)^2+i\varepsilon]
  (p^2-m^2+i\eps)}.
\end{equation}
The factor $N_{\delta D}$ is
\begin{eqnarray}
\label{eq:NdeltaD}
N_{\delta D}
&\equiv&
\gamma^\mu(\slashed{k}+m)\gamma^\nu
  \, 
  \frac{
  \bar{n}_\mu (p-k)_\nu + (p-k)_\mu \bar{n}_\nu}{p^+-k^+}\,
  \nonumber \\
  &=&
  \frac{2k^+\slashed{p}_\perp - 2p^+ \slashed{k}_\perp}{p^+-k^+}
  +2m
  +
  \frac{2k^+(p^- - k^-)\slashed{\bar{n}}+ 2(k_\perp \cdot p_\perp - k_\perp^2)  \slashed{\bar{n}}}{p^+-k^+}.
\end{eqnarray}
We note that the first two terms are of $\mathcal{O}(\lambda)$,
while the third term is of $\mathcal{O}(\lambda^2)$. A key feature of
      this expression is the appearance of the factor 
$1/(p^+-k^+)$, which originates from the additional pole in the
      gauge terms $\delta D^{\mu\nu}$ of the light-cone-gauge propagator.
From the fact that $i{\cal A}_{\rm QCD}^{\delta D}$ is nonzero, we
      see that the fermion self-energy is, as expected, gauge dependent.

\subsection{SCET calculations of the quark self-energy in the
        light-cone gauge}

Now let us compute the contributions to the quark self energy in the
light-cone gauge in the direct-QCD and modified building-block SCET
bases.  As we have already remarked, the light-cone-gauge calculations
are identical in these two bases. Hence, we use the collinear functions
$E$ [Eq.~(\ref{eq:collinear-fn-E})] and $J$
[Eq.~(\ref{eq:collinear-fn-J})] that are associated with the direct-QCD
building-block basis. We use the Feynman rules that derive from the
light-cone-gauge expressions for the bases in Eq.~(\ref{eq:bb-lc}),
which are identical to those for the direct-QCD basis.

First, let us consider the contributions of diagram (a) in
      Fig.~\ref{fig:SCETself} to the collinear functions $E$ and
      $J$. For this diagram, the Feynman rule for the quark-gluon vertex
      is unaffected by the light-cone-gauge condition $\bar n\cdot
      G=0$. Hence, we can split the contributions from this diagram 
 according to
\begin{align}
\label{eq:E-lc}
E^\text{LC}(u_1,u_2)\big|_{\textrm{(a)}}
&= E^\text{FG}(u_1,u_2)\big|_{\textrm{(a)}}
   + \delta E(u_1,u_2)\big|_{\textrm{(a)}}, \\[4pt]
\label{eq:J-lc}
J^\text{LC}(u_1,u_2)\big|_{\textrm{(a)}}
&= J^\text{FG}(u_1,u_2)\big|_{\textrm{(a)}}
   + \delta J(u_1,u_2)\big|_{\textrm{(a)}},
\end{align}
use our previous Feynman-gauge results, and compute $\delta E$ and
$\delta J$ directly.  Then, we have
\begin{align}
\label{eq:deltaE}
\delta E(u_1,u_2)\big|_{\textrm{(a)}}
&= ig_s^2 C_F\,
   \delta(u_1-\tfrac{1}{2})\delta(u_2-\tfrac{1}{2})
   \int_k
   \frac{P_n\,v(p)\bar{u}(p)\;
         N_{\delta D}\;
         (\slashed{p}+m)P_{\nbar}}
        {(k^2-m^2+i\eps)(p^2-m^2+i\eps)[(p-k)^2+i\eps]},
\end{align}
where $N_{\delta D}$ is given in Eq.~\eqref{eq:NdeltaD}.  $\delta
      J|_{\textrm{(a)}}$ is equal to $\delta E$ times the right factor
      $[(\slashed{p}_\perp-m)/(-p^+)]\,\slnbar/2$, which arises from
      $\bar\phi_n$.

Next, let us consider the contribution from diagram (b) in
Fig.~\ref{fig:SCETself} to the collinear functions $E$ and $J$.  As in
the Feynman gauge, there is no contribution to $E$ from diagram (b). In
the Feynman gauge, a contribution to $J$ arises from the gluon field
$\bar{n}\cdot G_n$ in the denominator of $\bar\phi_n$ in
Eq.~(\ref{eq:building-blocks-1d}).  This contribution vanishes in the
light-cone gauge.  Therefore, we cannot compute $\delta
J\vert_{\textrm{(b)}}$ directly, but must, instead, compute $J^{\rm
  LC}\vert_{\textrm{(b)}}$ and subtract our previous result for $J^{\rm
  FG}\vert_{\textrm{(b)}}$ to obtain $\delta
J\vert_{\textrm{(b)}}$. $J^{\rm LC}\vert_{\textrm{(b)}}$ receives a
contribution that arises from the transverse-gluon field in $\bar\phi_n$
in Eq.~(\ref{eq:phibar-lc}):
\begin{align}
\label{eq:Jb-LC}
J^\text{LC}(u_1,u_2)\big|_\text{(b)}
&= ig_s^2C_F\,\delta(u_1-\tfrac{1}{2})\delta(u_2-\tfrac{1}{2})
\int_k
   \frac{P_n v(p)\bar{u}(p)\,\gamma^{\mu}(\slashed{k}+m)
         P_{\bar{n}}\frac{\gamma_\perp^\nu}{-p^+}\frac{\slashed{\bar{n}}}{2}}
        {(k^2-m^2+i\eps)[(p-k)^2+i\eps]}
\nonumber \\
&
\times 
\left(
  g_{\mu\nu}
  - \frac{\bar{n}_\mu (p-k)_\nu + (p-k)_\mu \bar{n}_\nu}{\bar{n}\cdot (p-k)}
\right)
\nonumber 
\\
&=
ig_s^2C_F\,\delta(u_1-\tfrac{1}{2})\delta(u_2-\tfrac{1}{2}) 
\int_k
\frac{
P_n 
v(p)\bar{u}(p)
\frac{1}{p^+}
N^{J_{\textrm{(b)}}^{\textrm{LC}}}
}
{(k^2-m^2+i\varepsilon)
[(p-k)^2+i\varepsilon]},
\end{align}
where
\begin{equation}
N^{J_{\textrm{(b)}}^{\textrm{LC}}}
=
(D-2)k^+\frac{\slashed{n}\slashed{\bar{n}}}{4}
-(D-2)m\frac{\slashed{\bar{n}}}{2}
+
\frac{\left[(D-4)p^+-(D-2)k^+\right]}{p^+-k^+}
\slashed{k}_\perp \frac{\slashed{\bar{n}}}{2}
+
\frac{2 k^+}{p^+-k^+}
\slashed{p}_\perp \frac{\slashed{\bar{n}}}{2}.
\end{equation}

By making use of the results in Eqs.~\eqref{eq:delta_QCD},
      \eqref{eq:deltaE}, \eqref{eq:Jb-LC}, and \eqref{eq:JB-in-Feynman},
      we find that the direct-QCD and modified building-block bases in
      the light-cone gauge reproduce the full-QCD result in the
      light-cone gauge:
\begin{equation}
i\delta\mathcal{A}_{E_{\textrm{(a)}}}
+ i\delta\mathcal{A}_{J_{\textrm{(a)}}}
+ i\mathcal{A}_{J_{\textrm{(b)}}}^\text{LC}
         -i \mathcal{A}_{J_{\textrm{(b)}}}^\text{FG}
= i\mathcal{A}_\text{QCD}^{\delta D}.
\end{equation}

\subsection{Correspondence between gauge terms and Wilson lines
\label{sec:correspondence}}

Here we demonstrate the correspondence between the contributions to
      the quark self energy that arise from the gauge terms in the
      light-cone-gauge propagator and Wilson-line contributions in the
      Feynman gauge. Consider the gauge-term contribution $i{\cal
        A}_{\rm QCD}^{\delta D}$ in Eq.~(\ref{eq:delta_QCD}).  We write
      it as
\begin{eqnarray}
i{\cal A}_{\rm QCD}^{\delta D}
&=& 
ig_s^2 C_F \text{Tr} \int_k v(p)\bar u(p)
  \bigg[\frac{(\slashed{k}+\slashed{p}-m)-(\slashed{p}-m)}{k^+}
\frac{1}{\slashed{k}+\slashed{p}-m+i\varepsilon}\slashed{\bar n}
\nonumber\\
&&+\slashed{\bar n} \frac{1}{-\slashed{k}+\slashed{p}-m+i\varepsilon}
\frac{(-\slashed{k}+\slashed{p}-m)-(\slashed{p}-m)}{-k^+}\bigg]
\frac{1}{\slashed{p}-m+i\varepsilon}
\frac{H_{\textrm{QCD}}}{k^2+i\varepsilon},
\end{eqnarray}
where we have made the change of variables $k\to k+p$, so that $k$ is
now the gluon momentum, and we have made the further change of variable
$k\to -k$ in the second of the two main terms in square brackets.
The numerator terms $(\slashed{k}+\slashed{p}-m)$ and
$(-\slashed{k}+\slashed{p}-m)$ each cancel their respective denominator
factors, and the results cancel each other.\footnote{In fact, these
contributions are zero individually because of the antisymmetry of the
integrand under $k\to -k$.  However, we show the graphical-Ward-identity
cancellation, which does not rely on the (anti)symmetry of the
integrand.} The numerator factor $(\slashed{p}-m)$ in the first main
term in square brackets vanishes on acting on the spinor $\bar u(p)$,
while the numerator factor $(\slashed{p}-m)$ in the second main term in
square brackets cancels the corresponding denominator factor.  The
result, after taking $k\to -k$, is
\begin{eqnarray}
i{\cal A}_{\rm QCD}^{\delta D}&=& -ig_s^2 C_F \text{Tr} \int_k v(p)\bar u(p)
  \slashed{\bar n} \frac{1}{\slashed{k}+\slashed{p}-m+i\varepsilon}
\frac{1}{k^++i\varepsilon} \frac{H_{\textrm{QCD}}}{k^2+i\varepsilon},
\end{eqnarray}
which we recognize as the Wilson-line contribution that is shown in
Fig.~\ref{fig:two-point}(b). Here, we have inserted an
$i\varepsilon$ term into the denominator $k^+$ in order to define the
singularity at $k^+=0$ in accordance with the Wilson-line definition.

In a similar fashion, we can write the contribution to the two-point
function of the gauge terms in the light-cone-gauge propagator [$i{\cal
        A}_{\rm QCD}^{\delta D}$ in Eq.~(\ref{eq:delta_QCD})] as
\begin{eqnarray}
i\bar G_{\rm QCD}^{\delta D}&=& -g_s^2C_F\int_k \frac{1}{\slashed{p}-m+i\varepsilon}
  \bigg[\frac{(\slashed{k}+\slashed{p}-m)-(\slashed{p}-m)}{k^+}
\frac{1}{\slashed{k}+\slashed{p}-m+i\varepsilon}\slashed{\bar n}
\nonumber\\
&&+\slashed{\bar n} \frac{1}{-\slashed{k}+\slashed{p}-m+i\varepsilon}
\frac{(-\slashed{k}+\slashed{p}-m)-(\slashed{p}-m)}{-k^+}\bigg]
\frac{1}{\slashed{p}-m+i\varepsilon}
\frac{1}{k^2+i\varepsilon}.
\end{eqnarray}
Now, the factor $(\slashed{p}-m)$ in the first main term in the square
brackets cancels the corresponding denominator on the left, rather than
acting on a spinor.  The result, after the same cancellations as in the
case of $i{\cal A}_{\rm QCD}^{\delta D}$, is 
\begin{eqnarray}
i\bar G_{\rm QCD}^{\delta D}&=&g_s^2 C_F
\int_k
 \bigg(\frac{1}{\slashed{k}+\slashed{p}-m+i\varepsilon}
\slashed{\bar n}
\frac{1}{\slashed{p}-m+i\varepsilon}\frac{1}{k^+-i\varepsilon}\nonumber\\
&&\qquad+\frac{1}{\slashed{p}-m+i\varepsilon}\slashed{\bar n} 
\frac{1}{\slashed{k}+\slashed{p}-m+i\varepsilon}
\frac{1}{k^++i\varepsilon}\bigg)
\frac{1}{k^2+i\varepsilon}.
\end{eqnarray}
We recognize the first term in parentheses as the Wilson-line
contribution of Fig.~\ref{fig:two-point}(c) in the Feynman gauge
and the second term in
parentheses as the Wilson-line contribution of
Fig.~\ref{fig:two-point}(b) in the Feynman gauge.  Again, we
have inserted $i\varepsilon$ terms into the denominators $k^+$ in order
to define the singularities at $k^+=0$ in accordance with the
Wilson-line definition.

%%%%%%%%%%%%%%%%%%%%%%%%%%%%%%%%%%%%%1%%%%%%%%%%%%%%%%%%%%%%%%%


\begin{thebibliography}{999}

%\cite{Bauer:2000ew}
\bibitem{Bauer:2000ew}
C.~W.~Bauer, S.~Fleming and M.~E.~Luke,
Summing Sudakov logarithms in $B \to  X_s \gamma $ in effective field theory,
\href{http://dx.doi.org/10.1103/PhysRevD.63.014006}
{Phys. Rev. D \textbf{63}, 014006 (2000)}
[\href{https://arxiv.org/abs/hep-ph/0005275v1}{arXiv:hep-ph/0005275 [hep-ph]}].
%1009 citations counted in INSPIRE as of 05 Nov 2022

%\cite{Bauer:2000yr}
\bibitem{Bauer:2000yr}
C.~W.~Bauer, S.~Fleming, D.~Pirjol and I.~W.~Stewart,
An Effective field theory for collinear and soft gluons: Heavy to light decays,
\href{http://dx.doi.org/10.1103/PhysRevD.63.114020}
{Phys. Rev. D \textbf{63}, 114020 (2001)}
[\href{https://arxiv.org/abs/hep-ph/0011336v3}{arXiv:hep-ph/0011336 [hep-ph]}].
%1606 citations counted in INSPIRE as of 05 Nov 2022

%\cite{Bauer:2001ct}
\bibitem{Bauer:2001ct}
C.~W.~Bauer and I.~W.~Stewart,
Invariant operators in collinear effective theory,
\href{http://dx.doi.org/10.1016/S0370-2693(01)00902-9}
{Phys. Lett. B \textbf{516}, 134-142 (2001)}
[\href{https://arxiv.org/abs/hep-ph/0107001}{arXiv:hep-ph/0107001 [hep-ph]}].
%828 citations counted in INSPIRE as of 05 Nov 2022

%\cite{Bauer:2001yt}
\bibitem{Bauer:2001yt}
C.~W.~Bauer, D.~Pirjol and I.~W.~Stewart,
Soft collinear factorization in effective field theory,
\href{http://dx.doi.org/10.1103/PhysRevD.65.054022}
{Phys. Rev. D \textbf{65}, 054022 (2002)}
[\href{https://arxiv.org/abs/hep-ph/0109045}{arXiv:hep-ph/0109045 [hep-ph]}].
%1392 citations counted in INSPIRE as of 05 Nov 2022

%\cite{Bauer:2002nz}
\bibitem{Bauer:2002nz}
C.~W.~Bauer, S.~Fleming, D.~Pirjol, I.~Z.~Rothstein and I.~W.~Stewart,
Hard scattering factorization from effective field theory,
\href{http://dx.doi.org/10.1103/PhysRevD.66.014017}
{Phys. Rev. D \textbf{66}, 014017 (2002)}
[\href{https://arxiv.org/abs/hep-ph/0202088}{arXiv:hep-ph/0202088 [hep-ph]}].
%604 citations counted in INSPIRE as of 05 Nov 2022

%\cite{Beneke:2002ph}
\bibitem{Beneke:2002ph}
M.~Beneke, A.~P.~Chapovsky, M.~Diehl and T.~Feldmann,
Soft collinear effective theory and heavy to light currents beyond leading power,
\href{http://dx.doi.org/10.1016/S0550-3213(02)00687-9}
{Nucl. Phys. B \textbf{643}, 431-476 (2002)}
[\href{https://arxiv.org/abs/hep-ph/0206152}{arXiv:hep-ph/0206152 [hep-ph]}].
%741 citations counted in INSPIRE as of 05 Nov 2022

%\cite{Beneke:2002ni}
\bibitem{Beneke:2002ni}
M.~Beneke and T.~Feldmann,
Multipole expanded soft collinear effective theory with nonAbelian gauge symmetry,
\href{http://dx.doi.org/10.1016/S0370-2693(02)03204-5}
{Phys. Lett. B \textbf{553}, 267-276 (2003)}
[\href{https://arxiv.org/abs/hep-ph/0211358}{arXiv:hep-ph/0211358 [hep-ph]}].
%264 citations counted in INSPIRE as of 05 Nov 2022

%\cite{Pirjol:2002km}
\bibitem{Pirjol:2002km}
D.~Pirjol and I.~W.~Stewart,
A Complete basis for power suppressed collinear ultrasoft operators,
\href{http://dx.doi.org/10.1103/PhysRevD.69.019903}
{Phys. Rev. D \textbf{67}, 094005 (2003)}
[erratum: Phys. Rev. D \textbf{69}, 019903 (2004)]
[\href{https://arxiv.org/abs/hep-ph/0211251}{arXiv:hep-ph/0211251 [hep-ph]}].
%108 citations counted in INSPIRE as of 15 Dec 2022

%\cite{Bauer:2003mga}
\bibitem{Bauer:2003mga}
C.~W.~Bauer, D.~Pirjol and I.~W.~Stewart,
On Power suppressed operators and gauge invariance in SCET,
\href{http://dx.doi.org/10.1103/PhysRevD.68.034021}
{Phys. Rev. D \textbf{68}, 034021 (2003)}
[\href{https://arxiv.org/abs/hep-ph/0303156}{arXiv:hep-ph/0303156 [hep-ph]}].
%84 citations counted in INSPIRE as of 14 Dec 2022

%\cite{Lehmann:1954rq}
\bibitem{Lehmann:1954rq}
H.~Lehmann, K.~Symanzik and W.~Zimmermann,
On the formulation of quantized field theories,
\href{http://dx.doi.org/doi:10.1007/BF02731765}
{Nuovo Cim. \textbf{1}, 205-225 (1955)}
%574 citations counted in INSPIRE as of 14 Mar 2023

%\cite{Peskin:1995ev}
\bibitem{Peskin:1995ev}
M.~E.~Peskin and D.~V.~Schroeder,
``An Introduction to quantum field theory,''
Addison-Wesley, 1995,
ISBN 978-0-201-50397-5, 978-0-429-50355-9, 978-0-429-49417-8
doi:10.1201/9780429503559

%\cite{vanBijleveld:2025ekz}
\bibitem{vanBijleveld:2025ekz}
R.~van Bijleveld, E.~Laenen, C.~Marinissen, L.~Vernazza and G.~Wang,
Next-to-leading power jet functions in the small-mass limit in QED,
\href{http://dx.doi.org/doi:10.1007/JHEP07(2025)257}{JHEP \textbf{07}, 257 (2025)}
[\href{https://arxiv.org/abs/2503.10810}{arXiv:2503.10810 [hep-ph]}].
%3 citations counted in INSPIRE as of 14 Aug 2025

%\cite{Collins:1989gx}
\bibitem{Collins:1989gx}
J.~C.~Collins, D.~E.~Soper and G.~F.~Sterman,
Factorization of Hard Processes in QCD,
\href{http://dx.doi.org/10.1142/9789814503266\_0001}
{Adv. Ser. Direct. High Energy Phys. \textbf{5}, 1-91 (1989)}
[\href{https://arxiv.org/abs/hep-ph/0409313}{arXiv:hep-ph/0409313 [hep-ph]}].
%1514 citations counted in INSPIRE as of 15 Mar 2023

%\cite{Anikin:2009bf}
\bibitem{Anikin:2009bf}
I.~V.~Anikin, D.~Y.~Ivanov, B.~Pire, L.~Szymanowski and S.~Wallon,
QCD factorization of exclusive processes beyond leading twist: gamma*T ---\ensuremath{>} rho(T) impact factor with twist three accuracy,
\href{http://dx.doi.org/10.1016/j.nuclphysb.2009.10.022}{Nucl. Phys. B \textbf{828}, 1-68 (2010)}
[\href{https://arxiv.org/abs/0909.4090}{arXiv:0909.4090 [hep-ph]}].

%\cite{Marcantonini:2008qn}
\bibitem{Marcantonini:2008qn}
C.~Marcantonini and I.~W.~Stewart,
Reparameterization Invariant Collinear Operators,
\href{http://dx.doi.org/10.1103/PhysRevD.79.065028}{Phys. Rev. D \textbf{79} (2009), 065028}
[\href{https://arxiv.org/abs/0809.1093}{arXiv:0809.1093 [hep-ph]}].
%70 citations counted in INSPIRE as of 03 Dec 2025

%\cite{Beneke:2017ztn}
\bibitem{Beneke:2017ztn}
M.~Beneke, M.~Garny, R.~Szafron and J.~Wang,
Anomalous dimension of subleading-power N-jet operators,
\href{http://dx.doi.org/10.1007/JHEP03(2018)001}{JHEP \textbf{03}, 001 (2018)}
[\href{https://arxiv.org/abs/1712.04416}{arXiv:1712.04416 [hep-ph]}].
%58 citations counted in INSPIRE as of 28 Dec 2021

%\cite{Feige:2017zci}
\bibitem{Feige:2017zci}
I.~Feige, D.~W.~Kolodrubetz, I.~Moult and I.~W.~Stewart,
A Complete Basis of Helicity Operators for Subleading Factorization,
\href{http://dx.doi.org/10.1007/JHEP11(2017)142}
{JHEP \textbf{11}, 142 (2017)}
[\href{https://arxiv.org/abs/1703.03411}{arXiv:1703.03411 [hep-ph]}].
%65 citations counted in INSPIRE as of 16 Mar 2023

%\cite{Beneke:2018rbh}
\bibitem{Beneke:2018rbh}
M.~Beneke, M.~Garny, R.~Szafron and J.~Wang,
Anomalous dimension of subleading-power $N$-jet operators. Part II,
\href{http://dx.doi.org/10.1007/JHEP11(2018)112}
{JHEP \textbf{11}, 112 (2018)}
[\href{https://arxiv.org/abs/1808.04742}{arXiv:1808.04742 [hep-ph]}].
%38 citations counted in INSPIRE as of 17 Jan 2022

%\cite{Sterman:1993hfp}
\bibitem{Sterman:1993hfp}
G.~F.~Sterman,
``An Introduction to quantum field theory,''
Cambridge University Press, 1993,
ISBN 978-0-521-31132-8
%33 citations counted in INSPIRE as of 20 Dec 2025

%\cite{Fleming:2007xt}
\bibitem{Fleming:2007xt}
S.~Fleming, A.~H.~Hoang, S.~Mantry and I.~W.~Stewart,
Top Jets in the Peak Region: Factorization Analysis with NLL Resummation,
\href{https://journals.aps.org/prd/abstract/10.1103/PhysRevD.77.114003}{Phys. Rev. D \textbf{77} (2008), 114003}
%doi:10.1103/PhysRevD.77.114003
[\href{https://arxiv.org/pdf/0711.2079}{arXiv:0711.2079 [hep-ph]}].
%270 citations counted in INSPIRE as of 14 May 2026

\end{thebibliography}
\end{document}